\RequirePackage{snapshot}
\documentclass[lettersize,journal]{IEEEtran}

\usepackage{amsmath,amsfonts, amssymb, stmaryrd}
\usepackage{algorithmic}
\usepackage{algorithm}
\usepackage{array}
\usepackage{textcomp}
\usepackage{stfloats}
\usepackage{url}
\usepackage{verbatim}
\usepackage{graphicx}
\usepackage[table, xcdraw]{xcolor}
\usepackage{tikz}
\usepackage{subcaption}
\usepackage{tikzscale}
\usepackage{standalone}
\usepackage{xspace}
\usepackage{multirow}
\usepackage[numbers]{natbib}
\usepackage[backref=page]{hyperref}

\usetikzlibrary{backgrounds,fit,positioning,calc}

\newcommand{\A}{$\mathbf{A}$\xspace}
\newcommand{\B}{$\mathbf{B}$\xspace}
\newcommand{\C}{$\mathbf{C}$\xspace}
\newcommand{\D}{$\mathbf{D}$\xspace}
\newcommand{\E}{$\mathbf{E}$\xspace}
\newcommand{\F}{$\mathbf{F}$\xspace}
\newcommand{\Ap}{$\mathbf{A'}$\xspace}
\newcommand{\Bp}{$\mathbf{B'}$\xspace}
\newcommand{\Cp}{$\mathbf{C'}$\xspace}
\newcommand*\qf[1]{\textit{QF}_{\mathrm{#1}}}
\newcolumntype{C}[1]{>{\centering\let\newline\\\arraybackslash\hspace{0pt}}m{#1}}

\begin{document}

	\title{Dual JPEG Compatibility: a Reliable\\ and Explainable Tool for Image Forensics}
	\author{Etienne Levecque, Jan Butora, Patrick Bas\thanks{The authors are with the University of Lille, CNRS, Centrale Lille, UMR 9189 CRIStAL Lille, France. Email: etienne.levecque[at]gmail.com}}

	\markboth{}{}

	\maketitle

	\begin{abstract}
		Given a JPEG pipeline (compression or decompression), this paper demonstrates how to find the antecedent of an $8\times8$ block. If it exists, the block is considered \textit{compatible} with the pipeline. For unaltered images, all blocks remain compatible with the original pipeline; however, for manipulated images, this is not necessarily true. This article provides a first demonstration of the potential of compatibility-based approaches for JPEG image forensics. It introduces a method to address the key challenge of finding a block antecedent in a high-dimensional space, relying on a local search algorithm with restrictions on the search space.
		We show that inpainting, copy-move, and splicing, when applied after JPEG compression, result in three distinct mismatch problems that can be detected. In particular, if the image is re-compressed after modification, the manipulation can be detected when the quality factor of the second compression is higher than that of the first.
		Through extensive experiments, we highlight the potential of this compatibility attack under varying degrees of assumptions. While our approach shows promising results—outperforming three state-of-the-art deep learning models in an idealized setting—it remains a proof of concept rather than an off-the-shelf forensic tool. Notably, with a perfect knowledge of the JPEG pipeline, our method guarantees zero false alarms in block-by-block localization, given sufficient computational power.
	\end{abstract}


	\begin{figure}[!t]
		\centering
		\begin{subfigure}{0.48\columnwidth}
			\captionsetup{justification=centering}
			\includegraphics[width=\textwidth]{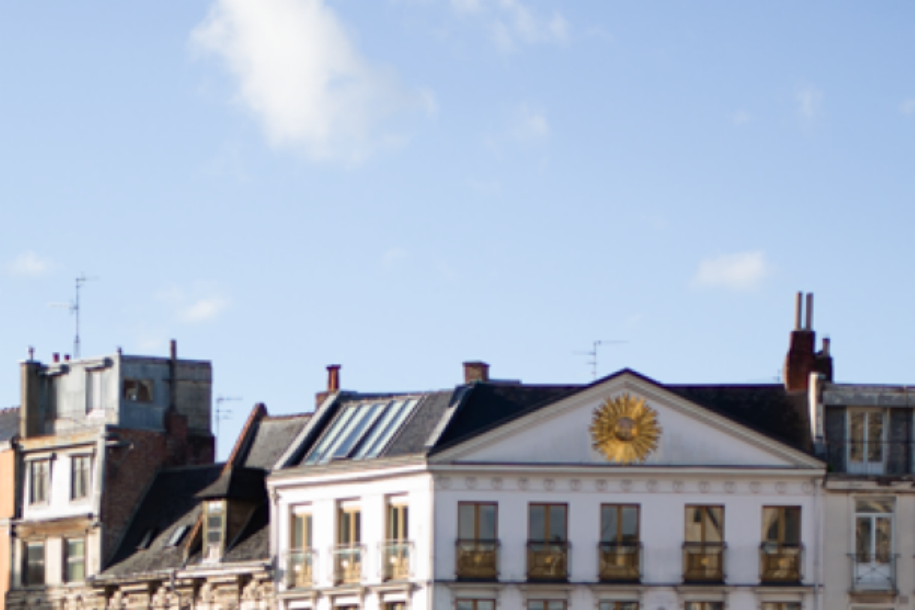}
			\caption{Original JPEG image at $\qf{1}=$ 80}
		\end{subfigure}
		\begin{subfigure}{0.48\columnwidth}
			\captionsetup{justification=centering}
			\includegraphics[width=\textwidth]{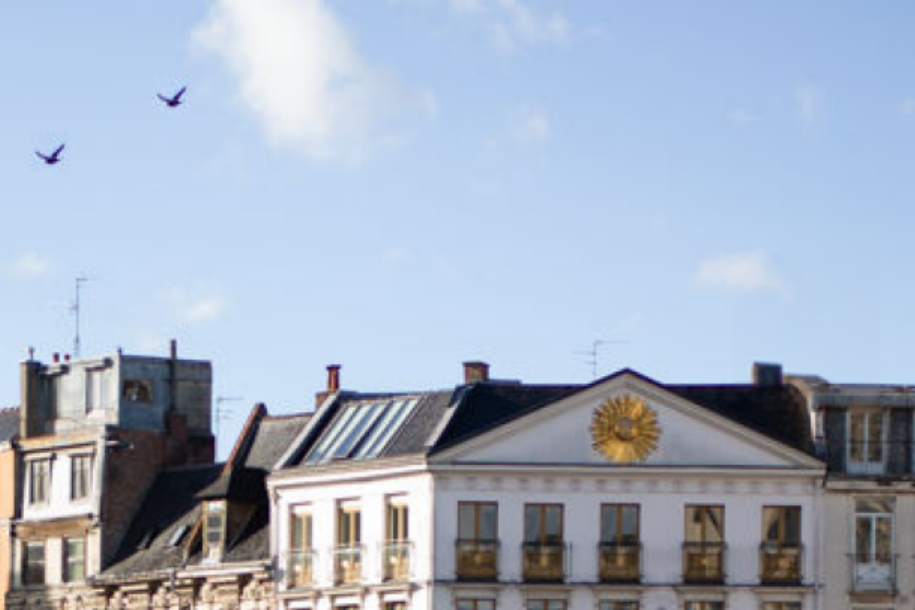}
			\caption{Forged JPEG image at $\qf{2}=$ 85}
		\end{subfigure}
		\begin{subfigure}{0.75\columnwidth}
			\captionsetup{justification=centering}
			\includegraphics[width=\textwidth]{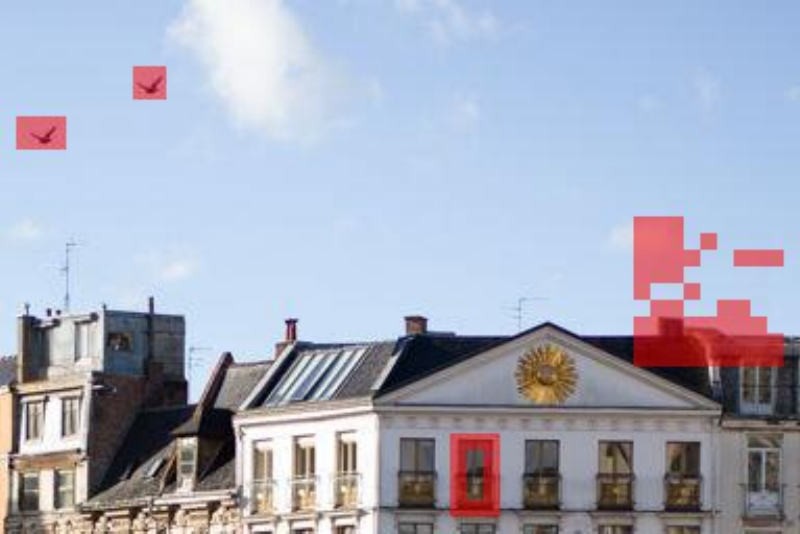}
			\caption{Detection results on the decompressed JPEG.}
		\end{subfigure}
		\caption{Demonstration of finding incompatible JPEG blocks. The chimney was removed using in-painting ({\it i.e.} uncompressed data), and birds were spliced with different Quality Factors ($\qf{}$s), one of them is fully aligned with the JPEG grid but not the other. The window was copy-moved aligned on the JPEG grid but not fully aligned, {\it i.e.} portions of blocks belong to the original image.}
		\label{fig:banner}
	\end{figure}

	The code related to this paper can be found on GitHub: \href{https://github.com/EtienneLevecque/jpeg-antecedent}{https://github.com/EtienneLevecque/jpeg-antecedent}.

	\section{Introduction}

	Nowadays, image editing has become incredibly easy, making reliable image forensic methods more crucial than ever. Almost anyone can alter an image using licensed software or free smartphone apps, and the results are often convincing. While most modifications are benign, some can become malicious. For instance, misinformation can involve deleting, adding, or altering important data in an image. Such images containing fake information can sometimes have serious consequences, including influencing public opinion in politics, falsifying scientific findings, or triggering harassment on social networks.

	In this paper, we leverage the fact that digital images are usually compressed. This is a common practice on the Internet, where reducing data traffic is a top priority. Of all the image compression methods, the most widespread is undoubtedly JPEG compression (Joint Photographic Experts Group). It was one of the first methods to be adapted to many applications and to retain good visual rendering. This popularity has been maintained over time, and many cameras still save images in JPEG by default. However, as we will see latter, many papers in the literature show that JPEG compression artifacts can be used to gather information about media authenticity.

	There are many different ways of altering an image. In our case, we are focusing on images that have already undergone JPEG compression, and were then modified in the pixel domain. The first modification is the {\it copy-move} which involves selecting a portion of the image, duplicating it, and pasting it elsewhere. If the portion comes from another image, this is known as {\it splicing}. Additionally, it is possible to locally change the pixel values to add or remove content, a process referred to as \textit{inpainting}.

	Our approach fits into this context of verifying the authenticity of JPEG images and, specifically, detecting local forgeries. The proposed method does not allow for the detection of global image forgeries or the generation of entire images. Our verification mechanism is based on the concept of block compatibility: if we observe an $8\times 8$ block \B, is there a block \A which, once passed through the JPEG pipeline, would give \B? If so, the block is compatible with the pipeline; if not, it is incompatible and has been altered.	By construction, all unmodified images are compatible with their development pipeline. But when modifications are made to the result of a JPEG pipeline, it can create a block that is impossible to obtain through the same pipeline: an incompatible block, and therefore a modified block. While the idea is simple, proving whether or not a block is compatible is a complex problem in very high dimensions. However, this approach has a significant advantage: unmodified images are inherently compatible, so there are no false positives.
	
	JPEG block compatibility has previously been applied in steganalysis~\cite{levecqueFindingIncompatiblesBlocks2024} to detect whether compressed images at $\qf{}100$ contained a hidden message. In that earlier work, the task of finding an antecedent was already addressed using a local search algorithm. However, this work differs in at least two directions from this previous work: it studied a different application of a similar algorithm but also improved this algorithm. 
	
	Regarding use cases, the steganalysis paper~\cite{levecqueFindingIncompatiblesBlocks2024} mainly addresses two questions: how to find antecedents of only DCT (Discrete Cosine Transform) blocks at a fixed $\qf{}100$ for a simple compression? And, what is the relation between manipulations in the DCT space and incompatible blocks at $\qf{}100$? In contrast, this paper tackles broader questions: how to find antecedents of pixel and DCT blocks for any $\qf{}$ across combined pipelines? And, what is the relation between manipulations in the pixel space, considering any $\qf{}$ and multiple compressions/decompressions, and incompatible blocks? 
	
	When it comes to algorithm improvement, there are three main differences. First, it now leverages an upper bound to reduce the search space. While this upper bound also existed in the steganalysis work, it was not useful to constrain the search space due to the specific quality factor. Second, after benchmarking different norms, we found that the infinity norm was more effective than the $\ell_1$ norm in the objective function. Third, the algorithm has been enhanced to handle any combination of compression and decompression, making it fully 'dual'—a capability that was not present in its initial version.

	\begin{figure*}[t]
		\centering
		\begin{tikzpicture}[node distance=\nodedistance]
			\def\nodedistance{1cm}
			
			\tikzstyle{arrow} = [thick,->,>=stealth]
			\tikzstyle{textbox} = [rectangle, align=center, draw=black, thick, minimum height=0.25*\nodedistance,minimum width=0.25*\nodedistance]
			
			\node[rectangle, align=center] (search) {Search antecedent\\$N$ iterations\\(\textit{cf.} Algorithm~\ref{alg:local_search})};
			\node[textbox, below=of search, yshift=0.75*\nodedistance] (neighborhood) {Define neighborhoord\\(\textit{cf.} Algorithm~\ref{alg:neighbors_function})\\using constraints $M_{\mathcal{F}}$\\(\textit{cf.} sec~\ref{sec:constraint})};
			
			\node[rectangle, align=center, draw=black, thick, fit=(search) (neighborhood)] (algorithm) {};
			\node[right=of algorithm] (result) {$\left\{
				\begin{aligned}
					&\text{Compatible}\\
					&\text{Incompatible}\\
					&\text{Unsolved}\\
				\end{aligned}\right.$};
			\node[textbox, left=of algorithm, yshift=0.8*\nodedistance] (image) {Image\\$\mathbf{D}$, $\mathbf{E}$, or $\mathbf{F}$\\(\textit{cf.} Fig.~\ref{fig:scenarios})};
			\node[textbox, left= of algorithm, yshift=-0.8*\nodedistance] (pipeline) {JPEG Pipeline\\$\mathcal{F}$\\(\textit{cf.} sec.~\ref{sec:JPEG pipeline})};
			
			\node[dashed,rounded corners=.1cm, draw, fit=(result) (algorithm) ] (blocks) {};
			\node at (blocks.north east)[fill=white, xshift=-1.85*\nodedistance] (test) {Repeat for each block};
			\node[textbox, right=of result] (output) {Reconstruct binary\\output mask};
			
			\draw[arrow] (image.east) -- (image.east -| algorithm.west);
			\draw[arrow] (pipeline.east) -- (pipeline.east -| algorithm.west);
			\draw[arrow] (algorithm.east) -- (result.west);
			\draw[arrow] (blocks.east) -- (output.west);
		\end{tikzpicture}
		\caption{General flowchart of our method. For each block of the image obtained with a pipeline $\mathcal{F}$, we search an antecedent with a local search constrained with upper bound $M_{\mathcal{F}}$. If the block is incompatible, we can conclude that it was altered. If the block is unsolved, the decision is left to the agent depending on the number of iterations $N$.}
		\label{fig:flowchart}
	\end{figure*}

	Our contributions are the following:
	\begin{itemize}
		\item We propose a dual extension of the antecedent search used in steganalysis~\cite{levecqueFindingIncompatiblesBlocks2024}. The previous method looks for a DCT antecedent of a pixel block, whereas the proposed algorithm can fit any pipeline with any number of compressions or decompressions. This search can find antecedents in the pixel or DCT domain and can find antecedents for integer DCT pipelines.
		\item A phylogenetic approach of all JPEG forgery scenarios (see Section~\ref{sec:compatibility}) shows that all pixel modification can fall into three types of compatibility mismatch: the grid mismatch, the quantization mismatch, and the pipeline mismatch. Solving each type of mismatch ensures solving the vast majority of JPEG forgery.
		\item We analyze the impact of the Quality Factor ($\qf{}$) over the compatibility result in each mismatch case and show that for some combinations of $\qf{}$ the search is fast on compatible blocks. Additionally, the detection is deterministic, \textit{i.e.} contrary to statistical methods, there are no false positives: a block detected as incompatible has been tampered with.
		\item Finally, we conduct experiments to show that compatibility can be used to build a JPEG forgery detector that achieves state-of-the-art performance. This method is very self-explainable, can gather information about the falsified areas, and localize forgery at the JPEG block level.
	\end{itemize}

	Fig.~\ref{fig:flowchart} illustrates the global strategy of the method presented in this paper. Section~\ref{sec:related work} presents works related to this paper, in particular about JPEG forgery detection and JPEG compatibility. The section~\ref{sec:JPEG pipeline} defines the main notations for JPEG compression, decompression, and composed pipelines. The compatibility is presented in section~\ref{sec:JPEGComp} along with the algorithm to prove that a block is compatible or not. Section~\ref{sec:compatibility} formulates all JPEG forgery scenarios into 3 mismatches that can create incompatible blocks and therefore could be detected with compatibility. Finally, section~\ref{sec:comparison} compares our JPEG forgery detector based on incompatibility with other JPEG-oriented detectors.
	
	\section{Related works}\label{sec:related work}
	
	To tackle the detection of falsified JPEG images, literature can be split into two categories: on one hand, there are the statistical methods built over an underlying constrained model: if the observed features do not follow the model, the image is classified as forged. On the other hand, there are statistical methods built with data that encompass supervised deep learning approaches that extract complex features, often lacking explainability, to classify or segment the image into falsified or non-falsified regions.

	\subsection{Forgery detection using statistical model}

	In the first category, Lukas {\it et al.}~\cite{lukasEstimationPrimaryQuantization2003} estimates the primary quantization table of a double-compressed image using the peaks in the DCT histograms. The main limitation is that it is hard to estimate the high-frequency quantization step, especially for low $\qf{}$s, because nearly all or all the coefficients are quantized to 0. 
	
	Lin {\it et al.}~\cite{lin2009fast} analyses the periodic pattern in the histograms of a DCT coefficient in the presence of double quantization. They show that in forged images with different regions (at least one being singly compressed and one being double compressed), it is possible to separate the blocks responsible for the periodic pattern from those that do not. This classification at the block level exhibits the localization of the forgery at the image level. 
	
	Farid {\it et al.}~\cite{faridExposingDigitalForgeries2009} propose a forgery detection based on a JPEG ghost. It is related to the convergence of multiple compression at the same quality factor: if a block was double compressed, compressing it at the same $\qf{}$ will return a very close value, but for an uncompressed block or block compressed with a different $\qf{}$, the difference will be more important. This difference reveals a darker region called "ghost". This method was automated for forgery detection in Zach's {\it et al.}~\cite{zachAutomatedImageForgery2012a} paper. Performances are good except when the modified area has been compressed with a lower $\qf{}$ than the $\qf{}$ of the image. 
	
	Luo {\it et al.}~\cite{luo2010jpeg} used the Laplacian model of AC DCT coefficients to build a statistic that can be used to identify if an image is JPEG compressed or not. It can also estimate the quantization steps or detect the quantization table in a dictionary. There is no result on forgery localization but it could be derived from the output of their method. 
	
	In Chen {\it et al.}~\cite{chen2011detecting}, the block artifact periodicity is used to determine if a block has been double compressed even after a grid shift. 
	
	Finally, Bianchi {\it et al.}~\cite{bianchi2012image} presents a more detailed model of the double quantized DCT coefficients in case of aligned or non-aligned double JPEG compression. Using an estimation of the first quantization table and the grid shift, a likelihood ratio test is used to classify each block as forged or authentic.
	
	Another category of methods relies on different versions of Benford's law~\cite{fu2007generalized,pasquini2014multiple,taimoriQuantizationUnawareDoubleJPEG2016} applied on DCT coefficients to derive statistical tests used to detect single/double compression or estimate the different $\qf{}$ used in the pipeline.
	
	\subsection{Forgery detection using deep learning}
	
	In the second category, we find deep learning detectors that specifically target modified JPEG images such as the convolutional neural networks proposed by Barni {\it et al.}~\cite{barniAlignedNonalignedDouble2017} to detect aligned or non-aligned double JPEG compression. Detecting double compression when the first quality factor is larger than the second one is a challenging task, but their results using the noise residuals and the pixel values are promising in this direction for non-aligned double compression. However, the generalization over multiple $\qf{}$s is not achieved since they need different datasets to train multiple models. 
	
	Cat-Net, the model proposed in Kwon's {\it et al.}~\cite{kwonLearningJPEGCompression2022a} paper takes these performances a step further with a good generalization over multiple forgery scenarios and $\qf{}$s. This model uses DCT volume representation instead of histograms and combines it with pixel values to detect JPEG compression artifacts at a pixel level.

	Also in the second category, there are deep learning models not specific to JPEG images such as ManTra-Net, the model proposed by Wu {\it et al.}~\cite{wuManTraNetManipulationTracing2019}. This model targets each type of image forgeries and the JPEG compression or double compression are only two classes among many others. If their visual segmentation is very satisfying, the misclassification error between single-compressed and double-compressed is close to 85\%. 
	
	Another general model is Noiseprint proposed by Cozzolino {\it et al.}~\cite{cozzolinoNoiseprintCNNBasedCamera2020} which relies on camera fingerprint estimation to detect that an image has been forged. Using contrastive learning, the authors train a model to estimate the so-called "noiseprint" of an image that can generalize this operation on unseen cameras. However, the main drawback of this scheme is its dependence on the $\qf{}$ since they need to train a specific detector for each quantization matrix. 
	
	TruFor is an improved version of Noiseprint proposed in Guillaro's {\it et al.} paper~\cite{guillaro2023trufor} which also relies on an estimation of camera noiseprint to localize forgeries. This model aims at detecting splicing but generalizes quite well on any forgery type when there is sufficient pattern in the fingerprint, which is the case with JPEG compression.
	
	Recently, SAFIRE has been presented in Kwon's {\it et al.} paper~\cite{kwon2024safire}. The main novelty of this model is the ability to do a multi-source segmentation of forged images. Moreover, the paper show very promising localization performances in comparison with others deep learning model, in particular, outperforming Cat-Net and TruFor.

	\subsection{JPEG compatibility}

	The compatibility notion in JPEG blocks is not recent, the first paper to present and use it, is the one of Fridrich {\it et al.}~\cite{fridrichSteganalysisBasedJPEG2001a}. To verify the compatibility of a block, they filter potential antecedent candidates with the $\ell_2$-distance from those candidates to the recompressed DCT block. They show that this distance can be upper-bounded by the norm of the worst-case scenario of rounding error. This constraint highly reduces the number of candidates for small $\qf{}$s but not enough for $\qf{}$s higher than 95. The method is very powerful but limited to grayscale images. The second limitation is that the guarantee for the compatibility is built with the mathematical Discrete Cosine Transform (DCT). Although the mathematical definition of the DCT is linear and bijective, most of the implementations do not respect those properties as detailed in section~\ref{sec:JPEGComp}.

	Compatibility has been extended to the reverse case (verifying the compatibility of a DCT block) in Butora's {\it et al.}~\cite{butora2019reverse} work. The authors propose a statistical model for the variance of the rounding error due to decompression. It is used to train a neural network to detect steganographic messages embedded in the DCT domain of JPEG images. A similar model was used in Dworetzky's {\it et al.}~\cite{dworetzky2023advancing} work to detect steganographic messages embedded in the spatial domain.

	The work of Lewis {\it et al.}~\cite{lewisExactJPEGRecompression2010} is not about detecting modification in JPEG images but doing exact recompression of a decompressed JPEG image. To do so, they try to find the exact DCT block of the observed pixel block, in other words, they search for an antecedent of a pixel block. The search is built with a very interesting set refining method: each step of the compressor tries to inverse a step of the decompressor by refining a set of possible values. At the end of the compression, the exact DCT antecedent should be the only one in the set. This method can deal with color images but the complexity at $\qf{}$s higher than 90 makes it intractable.

	An antecedent search was already used in~\cite{levecqueFindingIncompatiblesBlocks2024} to find pixel antecedents of a DCT block. The methodology is similar but was applied to steganalysis at $\qf{} = 100$. In the dual formulation formulated in the present paper (see section~\ref{sec:JPEGComp}), it appears that blocks are more likely to become incompatible when the modifications are done in the pixel domain which makes the JPEG compatibility much more general and allows forgery detection for a wide range of $\qf{}$s.

	\section{JPEG pipeline}\label{sec:JPEG pipeline}

	\subsection{Notations}

	We consider that an image is divided into non-overlapping blocks of size $(c, 8, 8)$ pixels with $c$ the number of channels (1 for grayscale, 3 for color images). The analysis of chrominance sub-sampling is beyond the scope of this study but should be addressed in a future work. We suppose for simplicity that the image size can be divided by 8. All operations in this paper are block independent, therefore we will use the same bold letter to refer to the image or to a block of this image. We use the letter $\mathbf{Q}$ to denote the quantization table of the same size $(c, 8, 8)$ as the block. The standard multiplication $x \times y$ and division $\frac{x}{y}$ symbols are used to define element-wise operations. The dot product $x\cdot y$ is used to refer to the matrix product. The rounding operator is defined as $\left[x\right]$. The notation tilde $\tilde{\cdot}$ is used to indicate a floating-point vector.\\

	We only present some mathematical formulation of the JPEG pipeline that will be used in the rest of the paper.

	\textbf{Color transformation}: Only RGB pixels undergo color transformation to become $YC_bC_r$ pixels (denoted YCC). The mathematical definition of this transformation is linear and bijective and can be applied independently for every pixel $\mathbf{x}_{\text{RGB}} \in [0;255]^3$ with the matrix $\mathbf{T}$ as follows:

	\begin{equation}
		\mathbf{x}_{\text{YCC}} = \mathbf{T} \cdot \mathbf{x}_{\text{RGB}} + \begin{pmatrix}
		0\\
		128 \\
		128
		\end{pmatrix}.
	\end{equation}
	\begin{equation}
	\mathbf{x}_{\text{RGB}} = \mathbf{T^{-1}} \cdot \left(\mathbf{x}_{\text{YCC}} - \begin{pmatrix}
	0\\
	128 \\
	128
	\end{pmatrix}\right).
	\end{equation}

	with $\mathbf{T} \cdot \mathbf{T}^{-1} = \mathbf{I}_3$ and,
	\begin{equation}
		\mathbf{T} = \begin{pmatrix}
		0.299&0.587&0.114\\
		-0.168736&-0.331264&0.5\\
		0.5&-0.418688&-0.081312
		\end{pmatrix}.
	\end{equation}

	However, most implementations of this color transformation round the result to the nearest integer and use only integer and bit shift (\texttt{libjpeg}~\cite{libjpeg} for instance) making them neither linear nor invertible.\\

	\textbf{JPEG compression}: The JPEG compression is applied independently on every channel of every block. Let $\mathbf{x}$ be a pixel block. Its Discrete Cosine Transform (DCT) coefficients $\mathbf{c}$ can be defined as follows:
	\begin{alignat}{2}
	\tilde{\mathbf{c}} &= \frac{\text{DCT}(\mathbf{x})}{\mathbf{Q}},\label{eq:def c_r}\\
	\mathbf{c} &= [\tilde{\mathbf{c}}].\label{eq:def c}
	\end{alignat}
	where $\text{DCT}$ refers to the forward 2D Discrete Cosine Transform function and $\mathbf{Q}$ is the quantization table of the same size as a block, which depends on the $\qf{}$ of the compression. A high $\qf{}$ means small quantization steps (at $\qf{}=100$, the quantization steps are all equal to $1$).
	DCT algorithm and rounding operation are not unique, there exist several types of implementation depending on the application. Therefore we use the notations $\mathbf{c} = \mathcal{C}(\mathbf{x}; \mathbf{Q})$ to define a specific compressor.\\

	\textbf{JPEG decompression}: The decompression process is almost symmetrical. Given a DCT block $\mathbf{c}$, the decompressed pixel block is:
	\begin{alignat}{2}
	&\tilde{\mathbf{y}} &&= \text{IDCT}(\mathbf{Q}\times \mathbf{c}),\label{eq:def x_r'}\\
	&\mathbf{y} &&= \left[\tilde{\mathbf{y}}\right]^{[0;255]},
	\label{eq:def x'}
	\end{alignat}
	where the notation $\left[\cdot\right]^{[0;255]}$ means that the results of the rounding operation are clipped to the set $[0;255]$. For the same reason as for the compressor, we use the notation $\mathbf{y} = \mathcal{D}(\mathbf{c}; \mathbf{Q})$ to define a specific implementation of a decompressor.\\

	As said for the compressor and decompressor, there exist several implementations of the DCT and IDCT algorithms to find a good trade-off between speed and precision for each application~\cite{loeffler1989practical, arai1988fast}. The same applies to the color transformation. In this paper, we focus on the default algorithm from \texttt{libjpeg} library~\cite{libjpeg} called \texttt{islow}. We also did all our experiments with the \texttt{libjpeg} color transformation that uses integers and thus is not lossless.

	\subsection{Composed Pipelines}\label{sec:composed pipeline}

	To deal with double compression or more means that compression and decompression must be composed multiple times with potentially different functions and parameters. We define those notations in this subsection.

	Let $f_1, \dots , f_n$ be $n>0$ compression or decompression functions alternated. This means that for each consecutive pair $(f_i, f_{i+1})$ there is a compression and a decompression. If $n=1$, the pipeline is only composed of a single function that can be either a compression or a decompression. Each function is parameterized by its quantization tables $\mathcal{Q} = (\mathbf{Q}_1, \dots, \mathbf{Q}_n)$. We denote the composed pipeline as $\mathcal{F}(\cdot\,; \mathcal{Q})$.

	For example, if we have $f_1$ a decompression with quantization matrix $\mathbf{Q}_1$ and $f_2$ a compression function with the quantization matrix $\mathbf{Q}_2$, then the composed pipeline of $f_1$ and $f_2$ is $\mathcal{F}$ parameterized with $\mathcal{Q} = (\mathbf{Q}_1, \mathbf{Q}_2)$ and for any block $\mathbf{x}$,

	\begin{equation}
	\mathcal{F}(\mathbf{x}; \mathcal{Q}) = f_2\left(f_1\left(\mathbf{x}; \mathbf{Q_1}\right); \mathbf{Q}_2\right).
	\end{equation}

	We also define the backward pipeline as $f_b$ ($\mathcal{F}_b$ for composed pipelines) that takes as argument a block from the same domain as $\mathbf{y}$ and return a block in the same domain as $\mathbf{x}$. For example, if $\mathcal{F}$ is a decompression using $\mathbf{Q}_1$ followed by a compression using $\mathbf{Q}_2$, $\mathcal{F}_b$ is a compression using $\mathbf{Q}_2$ followed by a decompression using $\mathbf{Q}_1$.

	Regardless of implementations used in the forward pipeline $\mathcal{F}$, the backward pipeline $\mathcal{F}_b$ can be defined with any DCT, IDCT, or color transformation algorithms. Therefore, we always define it with the floating-point bijective algorithms to avoid any error due to integer operation. Moreover, we do not round or clip any result in this backward pipeline, making it perfectly reversible.

	However, note that due to rounding operations, $f$ is not an invertible function, and thus $f$ and $f_b$ are not inverse functions of each other: $f\left(f_b\left(\mathbf{y}; \mathbf{Q}\right); \mathbf{Q}\right) \neq \mathbf{y}$ ($\mathbf{y}$ is an arbitrary block in the domain of $f_b$) and $f_b\left(f\left(\mathbf{x}; \mathbf{Q}\right); \mathbf{Q}\right) \neq \mathbf{x}$ most of the time. The same applies to $\mathcal{F}$ and $\mathcal{F}_b$.

	\section{JPEG Compatibility}\label{sec:JPEGComp}
	\subsection{Definitions and challenges}

	In this subsection, we define JPEG compatibility as the existence of a solution to an inverse problem.

	Let $\mathbb{S}$ be the set of pixel blocks or the set of DCT blocks. Let $\mathbf{y} \in \mathbb{S}$ be a block and let $f$ be a function of compression or decompression such that the codomain (set of destination) of $f$ is $\mathbb{S}$. Let $\mathbf{Q}$ be the quantization table that parameterizes $f$. We say that $\mathbf{y}$ is \textit{compatible} with $f$ if there exists a block antecedent $\mathbf{x}$ such that:

	\begin{equation}
	f\left(\mathbf{x}; \mathbf{Q}\right) = \mathbf{y}.
	\label{eq:inverse problem simple}
	\end{equation}

	On the contrary, if no antecedent $\mathbf{x}$ exists, we say that $\mathbf{y}$ is \textit{incompatible} with $f$.

	We can easily generalize this definition to any composed pipeline $\mathcal{F}$ with the set of quantization tables $\mathcal{Q}$ and codomain $\mathbb{S}$:
	\begin{equation}
	\mathcal{F}\left(\mathbf{x}; \mathcal{Q}\right) = \mathbf{y}.
	\label{eq:inverse problem composed}
	\end{equation}

	This inverse problem is complex because the pipeline is not an invertible function and thus applying the backward pipeline to $\mathbf{y}$ will return a close candidate but not always an antecedent. Even if we suppose that the uniform norm between this close candidate and the closest antecedent is 1, this means that for each element in the blocks, the error can be one of -1, 0, or 1. This sums to $3^{64}$ possibilities for grayscale blocks and $3^{192}$ possibilities for color blocks without colors sub-sampling. Brute forcing all solutions can not be considered, instead, we will rely on a local search to find antecedents combined with theoretical constraints to reduce the search space.

	\subsection{Local search to find antecedents}

	The local search is detailed in algorithm \ref{alg:local_search} and shares some similarities with the one presented in~\cite{levecqueFindingIncompatiblesBlocks2024}. However, this one can be used with any composed or simple pipeline and uses the $\ell_1$ norm instead of the infinity norm.

	\begin{algorithm}[t]
		\caption{Local search to find antecedent}
		\begin{algorithmic}
			\REQUIRE $\mathbf{y}, \mathcal{Q}$ \COMMENT{Observed block and quantization tables}
			\REQUIRE $\mathcal{F}$ \COMMENT{Pipeline}
			\REQUIRE $N > 0$ \COMMENT{Max iteration}
			\STATE $\mathbf{x}_s \gets \mathcal{F}_b(\mathbf{y})$ \COMMENT{Starting block of the search}
			\STATE add $\mathbf{x}_s$ to $P$ with cost $0$ \COMMENT{Priority queue initialization}
			\WHILE{$P$ not empty \AND $k \leq N$}
			\STATE $k \gets k + 1$
			\STATE $\mathbf{x} \gets$ remove first element of  $P$
			\FOR {$\mathbf{x}_n$ in $neighbors(\mathbf{x})$}
			\IF {$\mathbf{x}_n$ has not been visited}
			\IF {$\mathcal{F}(\mathbf{x}_n; \mathcal{Q}) = \mathbf{y}$}
			\RETURN $\mathbf{x}_n$ \COMMENT{$\mathbf{x}_n$ is an antecedent of $\mathbf{y}$}
			\ENDIF
			\STATE $c_n \gets \Vert \mathbf{y} - \mathcal{F}(\mathbf{x}_n; \mathcal{Q})\Vert_1$
			\STATE add $\mathbf{x}_n$ to $P$ with cost $c_n$ \COMMENT{$\mathbf{x}_n$ is a new candidate}
			\ENDIF
			\ENDFOR
			\ENDWHILE
		\end{algorithmic}
		\label{alg:local_search}
	\end{algorithm}

	The search begins by applying the backward pipeline to the observed block to obtain a starting candidate. Then, until the number of iterations is reached, we select the best candidate from a priority queue (queue data structure but every element has a priority; highest priority is always first), explore all its neighbors, and feed the priority queue again depending on the $\ell1$ distance to the observation. If the distance is 0, then an antecedent has been found and the search returns the result. A hash table is used to ensure that each block is visited only once.

	The neighborhood of a candidate is defined in Algorithm~\ref{alg:neighbors_function} with $M_{\mathcal{F}}$ an upper bound depending on the pipeline $\mathcal{F}$ that will be defined in the next subsection. However, when the block $\mathbf{x}$ is clipped, we ignore this upper bound and always add the new candidate to the neighborhood. Note that we only apply changes of $\pm1$ to one single value in the block but this process is repeated indefinitely until the maximum number of iterations is reached. Hence, every block in the constrained search space can be explored.

	\begin{algorithm}[t]
		\caption{Neighbors function}
		\begin{algorithmic}
			\REQUIRE $\mathbf{y}$ \COMMENT{Observed block}
			\REQUIRE $\mathbf{x}$ \COMMENT{Block for which a neighborhood is required}
			\REQUIRE $\tilde{\mathbf{x}}_b$ \COMMENT{Floating point starting point}
			\REQUIRE $M_{\mathcal{F}}$ \COMMENT{Upper bound of the pipeline}
			\IF {$\mathbf{y}$ is clipped}
			\STATE {$M_{\mathcal{F}} = \infty$}
			\ENDIF
			\FOR {$i$ in channels}
			\FOR {$j$ in rows}
			\FOR {$k$ in columns}
			\STATE $\mathbf{x}_n \gets \text{copy}(\mathbf{x})$
			\STATE $x_{n,i,j,k} \gets x_{n,i,j,k} + 1$ \COMMENT{Positive change}
			\IF {$\Vert\mathbf{x}_n - \tilde{\mathbf{x}}_b\Vert_2 \leq M_{\mathcal{F}}$}
			\STATE add $\mathbf{x}_n$ to the neighborhood
			\ENDIF
			\STATE $\mathbf{x}_n \gets \text{copy}(\mathbf{x})$
			\STATE $x_{n,i,j,k} \gets x_{n,i,j,k} - 1$ \COMMENT{Negative change}
			\IF {$\Vert\mathbf{x}_n - \tilde{\mathbf{x}}_b\Vert_2 \leq M_{\mathcal{F}}$}
			\STATE add $\mathbf{x}_n$ to the neighborhood
			\ENDIF
			\ENDFOR
			\ENDFOR
			\ENDFOR
			\RETURN neighborhood
		\end{algorithmic}
		\label{alg:neighbors_function}
	\end{algorithm}

	Note that to find an antecedent, this local search algorithm can be considered as fast. Indeed, the worst case is for $\qf{} = 100$ because lots of candidates are very close to the solution and can be solved in a matter of minutes for most blocks. However, the problem is for incompatible blocks, the search could explore the whole intractable space before emptying the priority queue. We can avoid that by constraining the searching space and thus the neighborhood.

	\subsection{Additional theoretical constraints}\label{sec:constraint}

	In order to speed up the search of incompatible blocks we also decided to see to which extent we can use the criterion proposed in the paper of Fridrich {\it et al.}~\cite{fridrichSteganalysisBasedJPEG2001a} where the authors defined a theoretical constraint for grayscale candidate blocks that drastically reduce the search space for $\qf{}< 95$. This constraint is presented in this subsection as well as the generalization to color images and double compression. All computations are done under two assumptions. First, we are using the mathematical definition of the DCT, the IDCT, and the color transformation such that all of them are unitary invertible functions. Second, we suppose that there is no clipping in the sense that all pixel values are always in the interval [0;255]. This second assumption is necessary otherwise, the rounding error can be bigger than $1/2$ and the constraints will not hold.

	Let's start with intermediate results that will simplify the equations. Let $\tilde{\mathbf{y}}$ be a floating point block of size $8\times8$ and $\mathbf{y} = [\tilde{\mathbf{y}}]$ its rounding value. We define the rounding error as $\mathbf{e} = \mathbf{y} - \tilde{\mathbf{y}}$. We have the following upper bound on the rounding error norm:
	\begin{equation}
	\label{eq:rounding error upper bound}
	\begin{aligned}
	\Vert\mathbf{y} - \tilde{\mathbf{y}} \Vert_2 &= \sqrt{\sum_{i=1}^{8} \sum_{j=1}^{8} \left\vert y_{i,j} - \tilde{y}_{i,j} \right\vert^2}\\
	&\leq \sqrt{\sum_{i=1}^{8} \sum_{j=1}^{8} \frac{1}{2}^2}\\
	&\leq 4.
	\end{aligned}
	\end{equation}

	Now if we have a color transformation, we use the underscript $RGB$ to denote the $3\times8\times8$ tensor with the three channels, and $Y$, $R$, $G$, and $B$ for each individual channel. The same result applies to other channels:
	\begin{equation}
	\begin{aligned}
	\left(T\cdot \mathbf{e}_{RGB}\right)_Y &= 0.299\mathbf{e}_R + 0.587\mathbf{e}_G + 0.114\mathbf{e}_B.
	\end{aligned}
	\end{equation}
	Since each row of $T$ sums to one, we can invoke the triangle inequality together with equation~\eqref{eq:rounding error upper bound} to upper bound each channel by $4$ and obtain:
	\begin{equation}
	\begin{aligned}
	\Vert T\cdot \left([\tilde{\mathbf{y}}_{RGB}] - \mathbf{y}_{RGB}\right)_Y \Vert_2 & \leq 4.
	\end{aligned}
	\end{equation}
	Note that we have the same upper bound when doing the inverse color transform.

	We now derive two upper bounds, one that is used when searching for an antecedent of a compression, and one used when searching antecedent of a decompression.

	Let us consider a simple compression setup with notations $\mathbf{x}, \mathbf{Q}, \tilde{\mathbf{c}}$ and $\mathbf{c}$ defined in equations \eqref{eq:def c_r} and \eqref{eq:def c}. This compression can be modeled by a pipeline $f$ and its backward equivalent $f_b$. We suppose that we observe $\mathbf{c}$ and would like to find the antecedent $\mathbf{x}$ from the starting point $\mathbf{x}_b = f_b(\mathbf{c})$. To speed up the search we will derive an upper bound to the distance between the true antecedent and the starting point:

	\begin{equation}
	\begin{aligned}
	\Vert \mathbf{x} - \mathbf{x}_b\Vert_2 & = \Vert \mathbf{x} - \text{IDCT}(\mathbf{Q}\times \mathbf{c})\Vert_2 \\
	& = \Vert \mathbf{x} - \text{IDCT}(\mathbf{Q}\times \mathbf{c} - \mathbf{Q}\times \tilde{\mathbf{c}} + \mathbf{Q}\times  \tilde{\mathbf{c}})\Vert_2 \\
	& = \Vert \mathbf{x} - \mathbf{x} - \text{IDCT}(\mathbf{Q}\times \mathbf{c} - \mathbf{Q}\times \tilde{\mathbf{c}})\Vert_2 \\
	& = \Vert \mathbf{Q}(\mathbf{c} - \tilde{\mathbf{c}}))\Vert_2 \\
	& \leq M_f = \left\Vert \frac{\mathbf{Q}}{2} \right\Vert_2
	\end{aligned}
	\end{equation}
	Indeed, the $\ell_2$-norm is invariant by unitary transformation (the IDCT) and the inequality comes from \eqref{eq:rounding error upper bound} with a quantization table.

	Now, let's consider a decompression setup with notations $\mathbf{Q}, \mathbf{c}, \tilde{\mathbf{y}}$ and $\mathbf{y}$ defined in equations \eqref{eq:def c}, \eqref{eq:def x_r'} and \eqref{eq:def x'}. This decompression can also be modeled by a pipeline $f$ and its backward equivalent $f_b$. We suppose that we observe $\mathbf{y}$ and would like to find the antecedent $\mathbf{c}$ from the starting point $\mathbf{c}_b = f_b(\mathbf{y})$. Using a very similar calculus, we derive an upper bound to the distance between the true antecedent and the starting point:

	\begin{equation}
	\begin{aligned}
	\Vert \mathbf{Q} \mathbf{c} - \mathbf{Q}\mathbf{c}_b\Vert_2 & = \Vert \mathbf{Q}\mathbf{c} - \text{DCT}(\mathbf{y})\Vert_2 \\
	& = \Vert \mathbf{Q} \mathbf{c} - \text{DCT}(\mathbf{y} - \tilde{\mathbf{y}} + \tilde{\mathbf{y}})\Vert_2 \\
	& = \Vert \mathbf{Q} \mathbf{c} - \mathbf{Q} \mathbf{c} - \text{DCT}(\mathbf{y} - \tilde{\mathbf{y}})\Vert_2 \\
	& = \Vert \mathbf{y} - \tilde{\mathbf{y}}\Vert_2 \\
	& \leq M_f = 4.
	\end{aligned}
	\end{equation}

	Using the triangle inequality of the norm, this constraint can be generalized to any pipeline by adding the upper bounds together. For example, if we want to search for an antecedent through a pipeline of decompression (upper bound is 4) with color transform (upper bound is also 4), the distance between the starting point and the true antecedent should not exceed $M_{\mathcal{F}} = 4 + 4$. In the double compression scenario, the pipeline is composed of a decompression ($M_f = 4$), a color transform ($M_f = 4$), a compression with quantization table $\mathbf{Q}$ ($M_f = \Vert \mathbf{Q}/2 \Vert_2$), another decompression ($M_f = 4$) and a final color transform ($M_f = 4$), then the distance should not exceed $M_{\mathcal{F}} = 4 + 4 + \Vert \mathbf{Q}/2 \Vert_2 + 4 + 4$.

	Finally, each upper bound obtained defines a sphere around the starting point in which the true antecedent should be. If we explore all candidates inside this sphere and do not find a solution to our inverse problem \eqref{eq:inverse problem simple}, then we can conclude that the "true antecedent" does not respect this property and the block is incompatible.
	
	\subsection{Analysis of the number of iterations}
	
	\begin{figure}[t]
		\centering
		\includegraphics[width=\columnwidth]{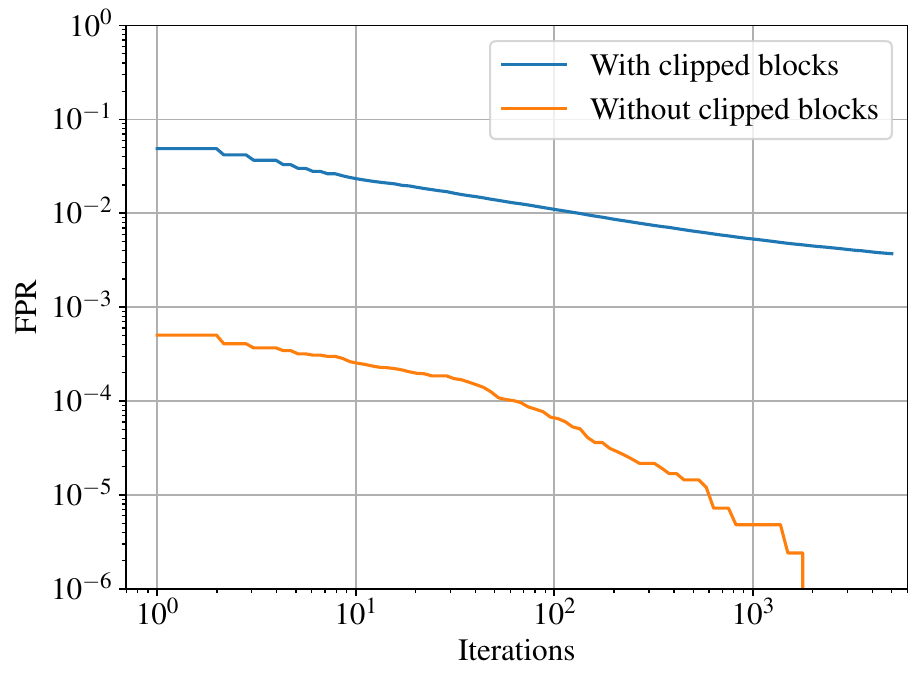}
		\caption{Probability for compatible blocks to be unsolved after a given number of iterations (FPR). This empirical curves were obtained by averaging the results of all single and double compression experiments ($\sim1$M blocks) from section~\ref{sec:comparison}.}
		\label{fig:iterations}
	\end{figure}
	
	The main parameter of the local search algorithm is the number of iterations. This subsection will present a short study of this parameter to help the reader select the correct value and see the influence of clipped blocks. Each block can be either compatible if an antecedent has been found, incompatible if no antecedent has been found among all possible candidates or unsolved if the algorithm reaches the maximum number of iteration before the two other cases. 
	
	Fig.~\ref{fig:iterations} highlights the impact of clipped blocks on the number of iterations. If we ignore them, the algorithm is able to find antecedent for all of the $\sim1$M compatible blocks in less that 2000 iterations and only 1 block in 5000 (equivalent to 1 block in an image of $560\times560$) is still unsolved after 1 iterations. However, clipped blocks are an issue because we can not use the reduction of the search space and we can see that even after 2000 iterations, there is still 1 block in 500 which is clipped and which is still unsolved.

	\begin{figure}[t]
		\centering
		\includegraphics[width=0.6\columnwidth]{scenarios.tex}
		\caption{General scenario for modifying a JPEG image in the spatial domain. Blues boxes represent blocks in the spatial domain (pixels) and orange boxes represent blocks in the frequency domain (DCT coefficients). $\mathcal{C}$ and $\mathcal{D}$ are the compressors and decompressors used. $\mathbf{Q}$ are the quantization tables. Modifications applied to $\mathbf{C}$ can be fully aligned (FA) or not fully aligned (NFA).}
		\label{fig:scenarios}
	\end{figure}
		
	\section{Compatibility of forged JPEGs}\label{sec:compatibility}

	We describe in this section the different types of forgery scenarios, their associated incompatibilities, and the results associated with each scenario.

	\subsection{Addressed forensics scenarios}

	There are several ways of forging a JPEG image and this section presents the underlying assumptions regarding the possible JPEG compression operations and the different scenarios that follow.

	In order to verify the JPEG compatibility of an image, we need at least a part of it to come from a decompressed JPEG image. We assume that there exists a pixel image \A, that has been compressed with a compressor $\mathcal{C}_1$ and quantization table $\mathbf{Q}_1$ to obtain the DCT coefficients \B. Once decompressed with a decompressor $\mathcal{D}_1$ using the same quantization table $\mathbf{Q}_1$, we obtain \C. This pixel image \C is the one that will be falsified.

	The general scenario, which includes all the others, is illustrated in Fig.~\ref{fig:scenarios}. In this graph, the double arrows towards \D mean that the image \D is built by using the values of the pixels from \C or \Cp and placing them at any position in \D. The other common part of each scenario relies on a second compression which can be applied on \D.

	Note that the notion of fully aligned or not fully aligned altering is inspired by the definitions of aligned and non-aligned recompression formalized by Bianci and Piva~\cite{bianchi2011detection}, and is illustrated in Fig.~\ref{fig:aligned}. A modification is \textit{fully aligned} (FA) if the JPEG grid of \C and the JPEG grid of \Cp are the same and if all modifications are done block-wise. For example, if a block of \D has some pixel values from C and some other pixel values from \Cp, then the modification is not full and therefore \textit{not fully aligned} (NFA) (\textit{cf.} Fig.~\ref{fig:aligned_c}). Note that the fully aligned situation is very rare when doing JPEG forgery. Indeed, if the grids can be aligned with a probability of $\frac{1}{64}$, blocks that form the boundary of the modification will most likely be cut, especially if a blur filter is applied to hide the contour artifacts.

	Three specific forensics scenarios are described as follows, each encompassing fully aligned or not fully aligned modifications.

	\textbf{Inpainting}. In this scenario, the modified pixel comes from a natural image \Cp that has not been compressed. Therefore, $\mathcal{C}_2$ and $\mathcal{D}_2$ do not exist and the modifications are necessarily non-aligned since \Cp is not associated with a JPEG grid.

	\textbf{Copy-Move}. In this scenario, \A=\xspace\Ap, \B=\xspace\Bp and \C=\xspace\Cp (compressors and decompressors are also equal) but the image \D is built by taking some pixels from \Cp and placing them somewhere else. The modifications can be aligned on the JPEG grid or non-aligned.

	\textbf{Splicing}. This is the most general scenario. Two images are compressed with potentially different compressors with some quantization tables $\mathbf{Q}_1$ and $\mathbf{Q}_3$ that can be equal or different. The decompressed versions of those images are used to build \D. The modifications can be aligned on the JPEG grid or non-aligned.

	\begin{figure}[t]
		\centering
		\begin{subfigure}{0.31\columnwidth}
			\captionsetup{justification=centering}
			\includegraphics[width=\textwidth]{aligned.tex}
			\caption{Fully aligned (FA)}
		\end{subfigure}
		\hfill
		\begin{subfigure}{0.31\columnwidth}
			\captionsetup{justification=centering}
			\includegraphics[width=\textwidth]{non_aligned_1.tex}
			\caption{Not fully aligned (NFA)}
			\label{fig:NFA_trans}
		\end{subfigure}
		\hfill
		\begin{subfigure}{0.31\columnwidth}
			\captionsetup{justification=centering}
			\includegraphics[width=\textwidth]{non_aligned_2.tex}
			\caption{Not fully aligned (NFA)}
			\label{fig:aligned_c}
		\end{subfigure}
		\caption{Examples of fully aligned and not fully aligned modifications. Black lines represent blocks. Example~\ref{fig:NFA_trans} illustrates block translations from one image to another, on the contrary example~\ref{fig:aligned_c} illustrates a finer segmentation.}
		\label{fig:aligned}
	\end{figure}
	
	\subsection{Possible sources of incompatibility}

	This subsection translates the forgery scenarios described in the previous one into three distinct mismatches that can create incompatible blocks. We denote them as the {\it grid} mismatch, the {\it quantization table} mismatch, and the {\it pipeline} mismatch.

	\textbf{Grid mismatch} appears when the dependency between the 64 values of a block (decompressed pixels or DCT coefficients) is broken. It can be broken either because the block was partially modified and therefore there is no dependency between the two parts of the block (NFA copy-move or NFA splicing) or because the modified pixels are simply not JPEG decompressed and thus not dependent (inpainting).

	\textbf{Quantization table mismatch} can only be observed when the forgery is fully aligned and when the quantization table used is not the same as the one used in the main pipeline. In this case, the 64 decompressed pixel values are dependent and are not consistent with the original pipeline. This mismatch can for example be obtained when doing fully aligned splicing.

\begin{figure}[t]
		\centering
		\includegraphics[width=0.6\columnwidth]{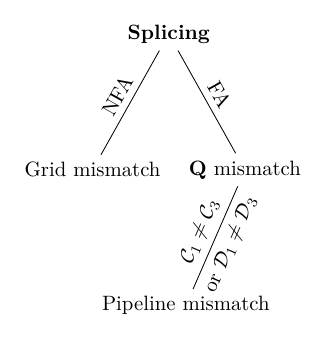}
		\caption{In the splicing scenario, if the modifications are not fully aligned (NFA), it is a \textit{grid} mismatch, in the fully aligned case (FA), there can be a \textit{quantization table} mismatch if tables are not the same. \textit{Pipeline} mismatch can appear if one of the compression or decompression (or both) differs. Note that a FA splicing with the same quantization tables and functions is a \textit{perfect} splicing in the sense that it does not create any mismatch.}
		\label{fig:splicing}
	\end{figure}

	\textbf{Pipeline mismatch} can also only be observed when doing fully aligned splicing. It comes from the fact that the functions used in the compression and decompression process can be different. For example, if the DCT transform does not have the same implementation or when the rounding function is not the same. It comes from any implementation difference that influences the actual result. In particular, Fig.~\ref{fig:splicing} illustrates the possible mismatches for the splicing scenario and shows that one can observe a single pipeline mismatch if compressors are not the same but decompressors are the same, or a double pipeline mismatch if both compressors and decompressors are different.
	
	In this classification, we assume there can only exist a single mismatch. Indeed, if there is a grid mismatch, it is impossible to define the quantization table, either because it does not exist at all or because it would mean that there are multiple quantization tables used in the image. For the same reason, the pipeline can not be correctly defined, therefore, the quantization table mismatch and the pipeline mismatch do not exist. Using the same reasoning, we assume that if the manipulation is fully aligned (there is no grid mismatch) then if a quantization mismatch exists we can not define the pipeline mismatch.

	In our forgery scenario, the agent performing the analysis can observe either \D, \E, or \F. Depending on the observed image and the type of mismatch, it may be necessary to search for an antecedent further up the pipeline. For example, to verify a grid mismatch while observing \E, we need to find a DCT antecedent $\mathbf{X}$ through the combination of $\mathcal{D}_1$ and $\mathcal{C}_2$ such that $\mathcal{C}_2(\mathcal{D}_1(\mathbf{X})) = \mathbf{E}$.
	
	\subsection{Statistics of incompatible blocks}

	In the previous section, three possible mismatches and three observations yielded nine different experiments. This section presents some experiments to explore the statistics of incompatible blocks and their relation to the $\qf{}$. We used images from the UCID dataset composed of 1338 RGB color images of size $384\times512$. These pixel images are equivalent to $\mathbf{A}$ in our general scenario (Fig.~\ref{fig:scenarios}). We use the \texttt{islow} DCT method from the \textit{libjpeg} library using all $\qf{}$ between 50 and 100 to create the compressed image $\mathbf{B}$. To obtain the decompressed pixel image $\mathbf{C}$ we use the decompressor of the same library. In this subsection, $\qf{1}$ and $\qf{2}$ refer to the true quality factors of a block and we also use them to denote the quantization table associated with this $\qf{}$. However, the tested $\hat{\qf{1}}$ refers to the $\qf{}$ used in the pipeline of the search.

	\begin{figure}[t]
		\centering
		\includegraphics[width=\columnwidth]{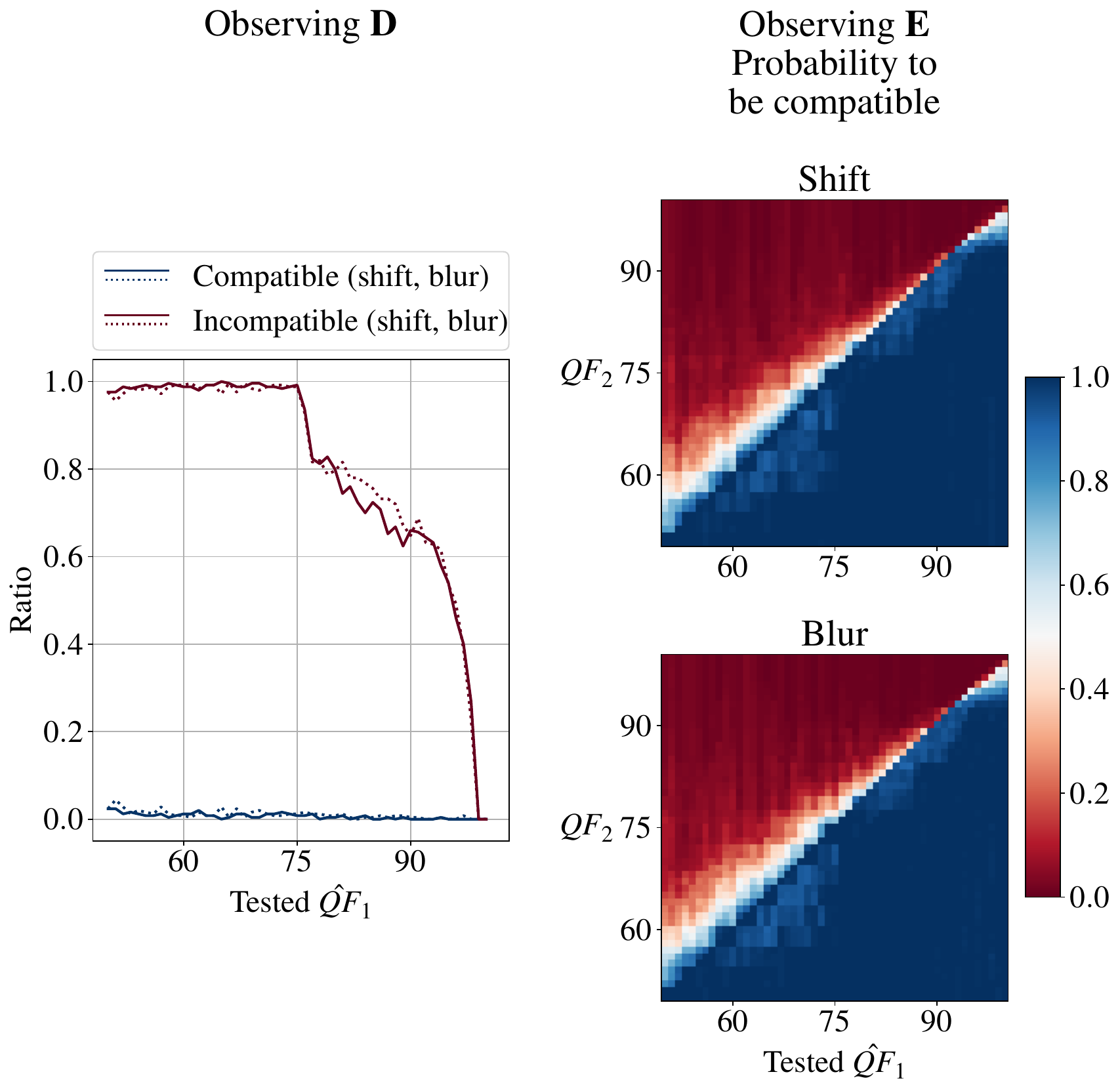}
		\caption{Statistics in the presence of a grid mismatch due to a grid shift or a blurring kernel to remove the grid. The left plot gives ratio of compatible (blue) and incompatible (red) blocks when observing $\mathbf{D}$ and the right plots when observing $\mathbf{E}$ (almost equal to plots when observing $\mathbf{F}$). Tested $\hat{\qf{1}}$ is the $\qf{}$ used to search an antecedent. Note that, all blocks have been modified in this experiment.}
		\label{fig:grid_mismatch_results}
	\end{figure}

	\begin{figure}[t]
		\centering
		\includegraphics[width=\columnwidth]{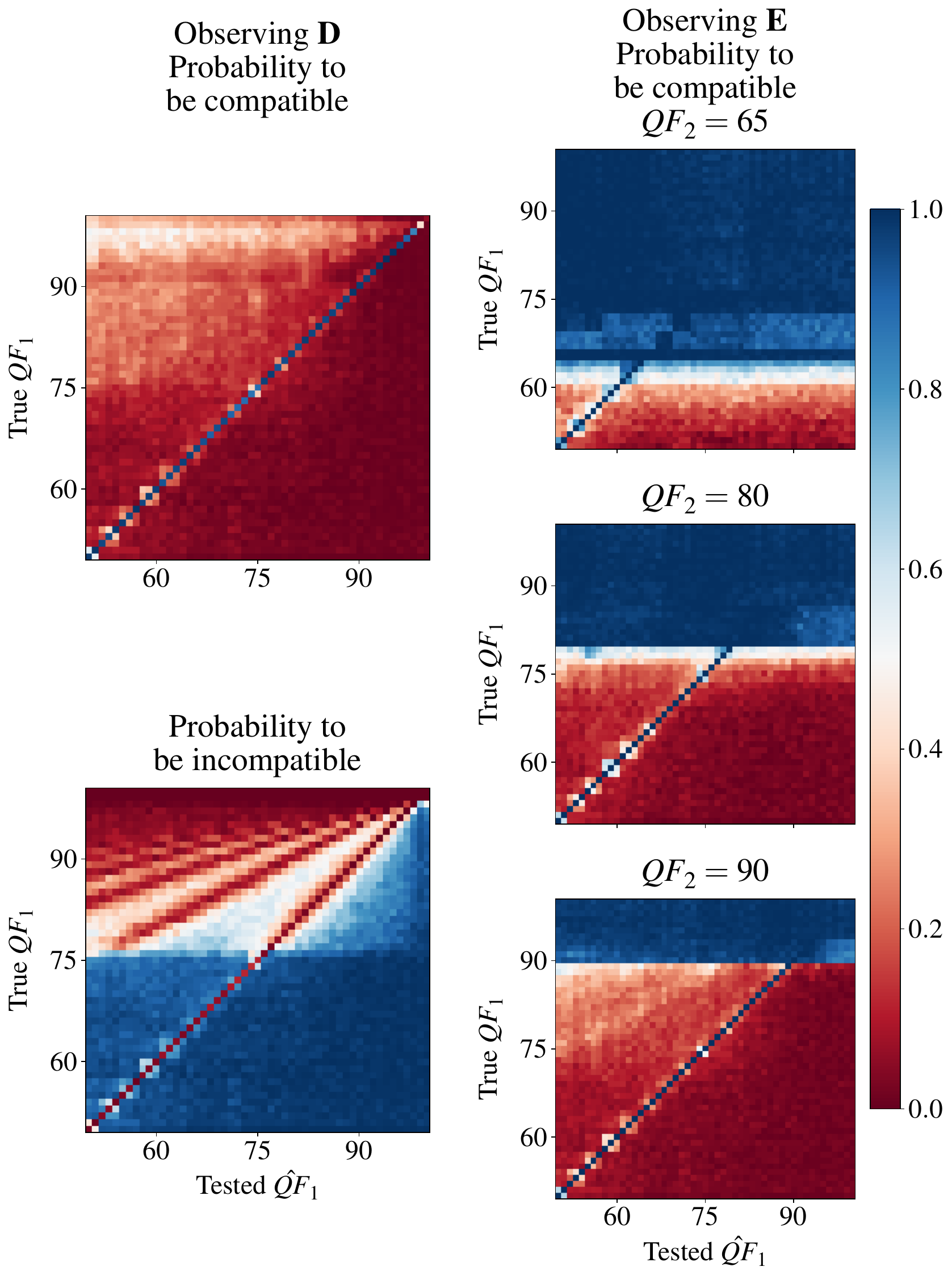}
		\caption{Statistics in the presence of a quantization mismatch. Left plots give statistics when observing $\mathbf{D}$ and right plots when observing $\mathbf{E}$ (almost equal to plots when observing $\mathbf{F}$). Tested $\hat{\qf{1}}$ is the $\qf{}$ used to search an antecedent.}
		\label{fig:q_mismatch_results}
	\end{figure}

	\textbf{Statistics on grid mismatch:} We have seen in the last subsection that grid mismatch can be caused by a grid shift (such as non-aligned splicing or copy-move) or by inpainting a region without a grid at all. To explore both cases, we try to find antecedents to blocks modified with a random misalignment (called shift) and blocks modified with a blurring kernel (called blur). We can draw the following conclusions by looking at the results in Fig.~\ref{fig:grid_mismatch_results}:
	\begin{itemize}
		\item When observing $\mathbf{D}$ ({\it i.e.} forged blocks in the pixel domain), blocks with a grid mismatch are very unlikely to be compatible with any tested $\hat{\qf{1}}$ (probability of being compatible almost equals to 0). For small enough tested $\hat{\qf{1}}$ we can even prove that blocks are incompatible (probability of being incompatible almost equals to 1). The search space becomes bigger with $\hat{\qf{1}}$, so it requires more iterations in the search Algorithm~\ref{alg:local_search} to explore all possible candidates. For this experiment the number of iterations was fixed to 100.
		\item When observing $\mathbf{E}$ ({\it i.e.} recompressed forged blocks), blocks with a grid mismatch are very unlikely to be compatible for tested $\qf{1}$ small enough compared to $\qf{2}$ (above diagonal tested $\hat{\qf{1}} = \qf{2}$).
		\item But when observing $\mathbf{E}$, if the tested $\hat{\qf{1}}$ is too big compared to $\qf{2}$, blocks with a grid mismatch are very likely to be compatible to this tested $\hat{\qf{1}}$ (below diagonal tested $\hat{\qf{1}} = \qf{2}$).
		\item Observing $\mathbf{E}$ or $\mathbf{F}$ gives very close results.
		\item Shift and blur yield the same probability, so an inpainted block has the same probability of being incompatible as a non-fully aligned spliced or copy-move block.
	\end{itemize}

	\textbf{Statistics on quantization mismatch:} This time, observed blocks are compatible to a true $\qf{1}$ but we want to evaluate how likely they are to be compatible to another tested $\hat{\qf{1}}$. Fig.~\ref{fig:q_mismatch_results} depicts the results of our experiments and again we can draw the following conclusions:
	\begin{itemize}
		\item When observing $\mathbf{D}$, blocks with a quantization mismatch are very unlikely to be compatible with any other tested $\hat{\qf{1}}$ different than the true $\qf{1}$. For small enough true $\qf{1}$ we can even prove that blocks are incompatible with any other pipeline using a different $\hat{\qf{1}}$. We observe the same separation as for simple compression around true $\qf{1} = 75$, which is due to the size of the search space compared to the maximum number of iterations in the search Algorithm~\ref{alg:local_search}.
		\item When observing $\mathbf{E}$, blocks with a quantization mismatch are very unlikely to be compatible with any other tested $\hat{\qf{1}}$ different than the true $\qf{1}$ if $\qf{1}$ is small enough compared to $\qf{2}$ (below the line true $\qf{1} = \qf{2}$).
		\item However, when observing $\mathbf{E}$, if $\qf{1}$ is bigger than $\qf{2}$, blocks with a quantization mismatch are very likely to be compatible with any tested $\hat{\qf{1}}$ (above the line true $\qf{1} = \qf{2}$).
		\item Observing $\mathbf{E}$ or $\mathbf{F}$ gives very close results.
	\end{itemize}
	
	Notice that in both Fig~\ref{fig:grid_mismatch_results} and Fig.~\ref{fig:q_mismatch_results}, when $\qf{2}$ is smaller than $\qf{1}$, all blocks become compatible. This occurs because a compression with a smaller $\qf{}$ acts as a projection in a smaller space (both in size and in dimension since some coefficient at set to 0). So if a block is incompatible because of a manipulation after the first compression, this second compression acts as a projection on a set of compatible blocks, and therefore, the block after the second compression is compatible. 
	
	\begin{figure*}
		\begin{subfigure}{0.16\textwidth}
			\includegraphics[width=\textwidth]{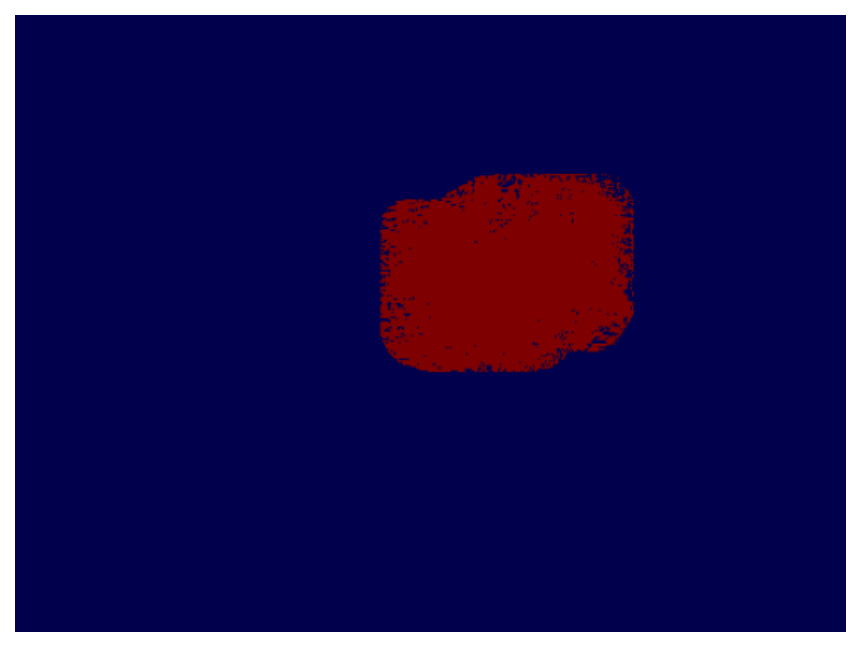}
			\caption{Mask}
			\label{fig:pipe_mismatch_mask}
		\end{subfigure}
		\begin{subfigure}{0.16\textwidth}
			\includegraphics[width=\textwidth]{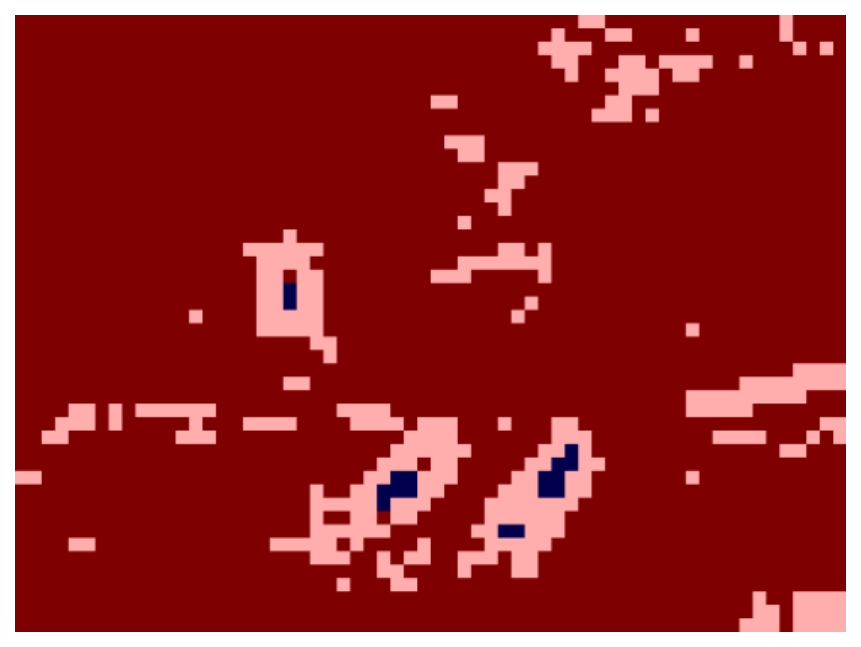}
			\caption{$\hat{\qf{1}} = 60$}
			\label{fig:pipe_mismatch_60}
		\end{subfigure}
		\begin{subfigure}{0.16\textwidth}
			\includegraphics[width=\textwidth]{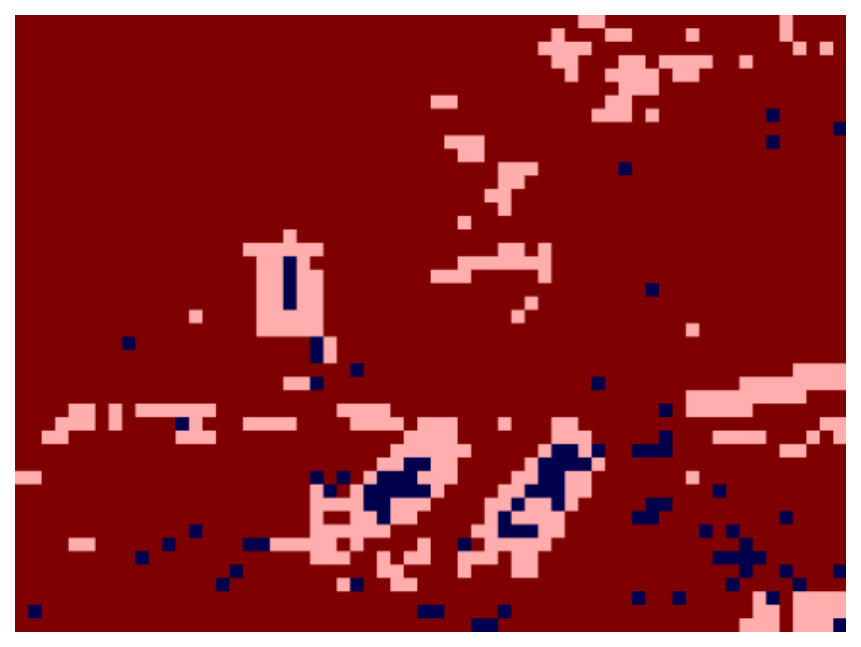}
			\caption{$\hat{\qf{1}} = 74$}
			\label{fig:pipe_mismatch_74}
		\end{subfigure}
		\begin{subfigure}{0.16\textwidth}
			\includegraphics[width=\textwidth]{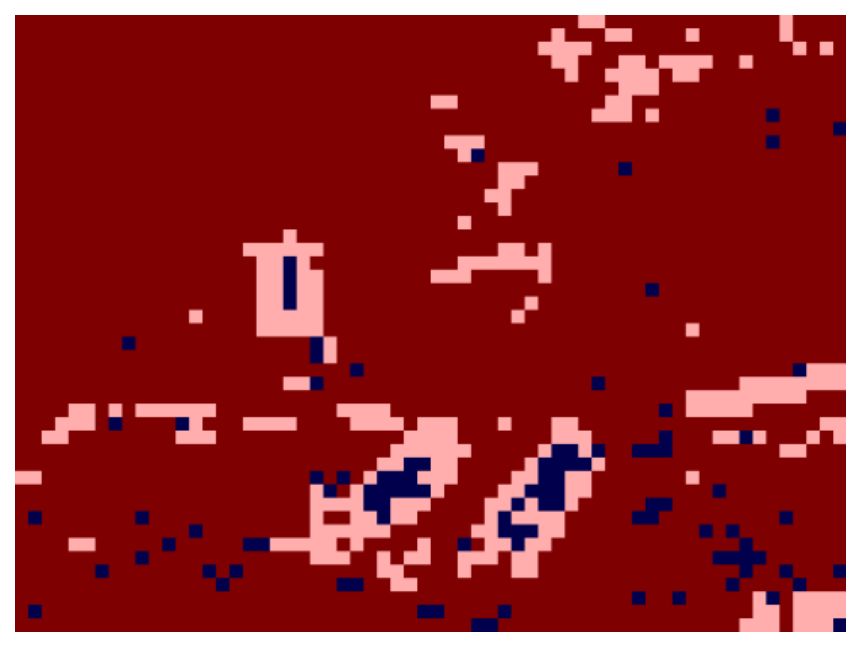}
			\caption{$\hat{\qf{1}} = 75$}
			\label{fig:pipe_mismatch_75}
		\end{subfigure}
		\begin{subfigure}{0.16\textwidth}
			\includegraphics[width=\textwidth]{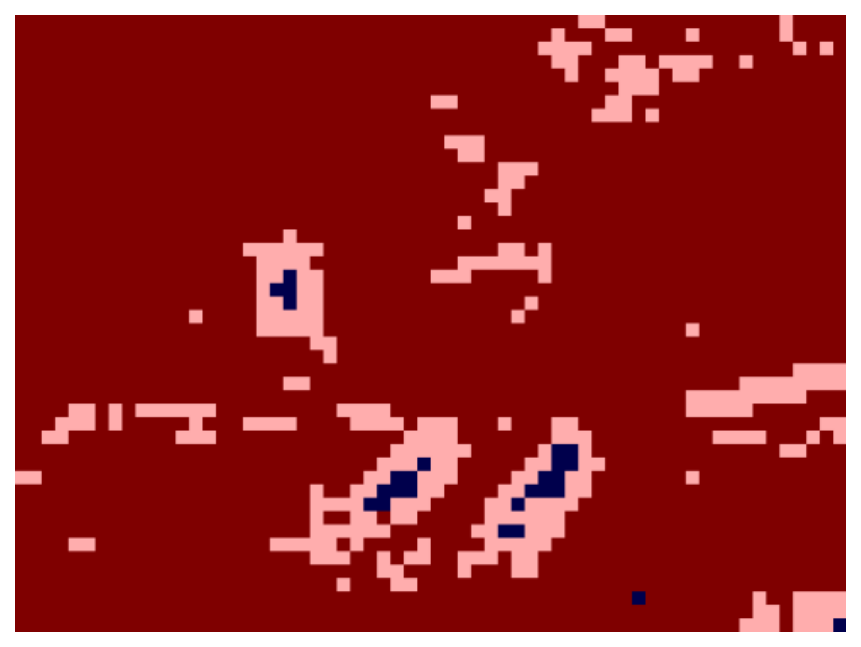}
			\caption{$\hat{\qf{1}} = 76$}
			\label{fig:pipe_mismatch_76}
		\end{subfigure}
		\begin{subfigure}{0.16\textwidth}
			\includegraphics[width=\textwidth]{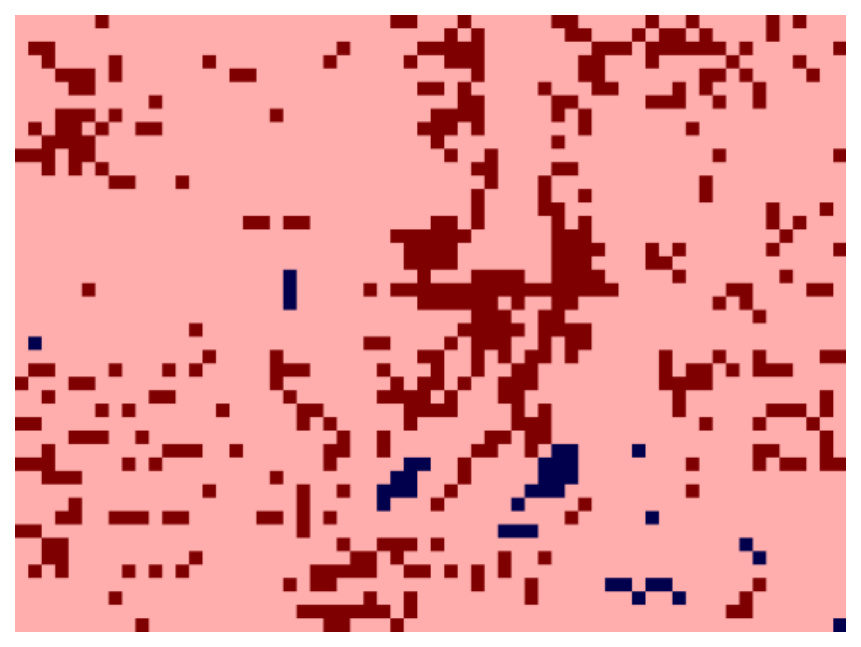}
			\caption{$\hat{\qf{1}} = 90$}
			\label{fig:pipe_mismatch_90}
		\end{subfigure}
		\caption{Result of the algorithm in the presence of a pipeline mismatch for a single compression after 5000 iterations. The original image was compressed with \texttt{islow} from \texttt{libjpeg} at $\qf{1} = 75$ but the algorithm used in the antecedent search is \texttt{float} from \texttt{libjpeg}. In red, are incompatible blocks, in pink unsolved blocks, and blue compatible blocks. The blue blocks in all the predictions are constant blocks that are always compatible with any DCT algorithm.}
		\label{fig:pipeline_mismatch}
	\end{figure*}

	\textbf{Pipeline mismatch:} Localizing a manipulation because of the pipeline mismatch is very unlikely to happen because this would imply that the manipulation is fully aligned and that there is no quantization mismatch, two rare scenarios. Nonetheless, it’s useful to explore this concept, especially when the pipeline is not fully known. For instance, consider the situation where we attempt to find the antecedent of compatible blocks using an incorrect DCT algorithm. We conduct a small experiment where an image is compressed at $\qf{1} = 75$ using the \texttt{islow} DCT algorithm from \texttt{libjpeg}. After the compression, this image is then manipulated. Next, we apply the antecedent search but with the \texttt{float} DCT algorithm also presented in \texttt{libjpeg} and for different $\hat{\qf{1}}$ used in the search. In Fig.~\ref{fig:pipeline_mismatch} we see that most blocks are classified as either incompatible or unsolved, with only a small fraction becoming compatible when $\hat{\qf{1}} = 75$, which is the correct $\qf{}$.  This experiment highlights how sensitive the algorithm is to variations in the pipeline. If the pipeline is not known, this experiment implies that it should be possible to do a dictionary attack using multiple pipelines and most of them will not work but the correct one.
	
	To sum up the statistics in this subsection, if we observe the image $\mathbf{D}$ which has been modified without recompression, there is a very high chance that modified blocks due to inpainting, splicing, or copy-move will be incompatible and detected as such (we assume that fully-aligned copy-move or fully-aligned splicing with the same quantization table are rare events). And if we observe $\mathbf{E}$ (or $\mathbf{F}$, the results should be almost the same) which have been recompressed with $\qf{2}$, we should be able to detect modified blocks due to inpainting, splicing or copy-move if $\qf{1}$ is smaller than $\qf{2}$. Otherwise, we will not be able to detect anything.

	\section{Forgery localization}\label{sec:comparison}
	
	\begin{figure*}[t]
		\centering
		\begin{tikzpicture}[node distance=1.9*\nodedistance]
			\def\nodedistance{1cm}
			
			\tikzstyle{arrow} = [thick,->,>=stealth]
			\tikzstyle{textbox} = [rectangle, align=center, draw=black, thick, minimum height=0.25*\nodedistance,minimum width=0.25*\nodedistance]
			
			\node[rectangle, align=center,draw=black] (flickr) {Flickr30k\\20 pxl img};
			\node[rectangle, right=of flickr, align=center,draw=black] (dct images) {$3\times 20$ \\DCT img};
			\node[rectangle, right=of dct images, align=center,draw=black, xshift=-0.8cm] (pixel images){$3\times 20$ \\pxl img};
			\node[rectangle, right=of pixel images, align=center,draw=black] (manipulated){$3 \times 20$ pxl img\\manipulated};
			\node[rectangle, right=of manipulated, align=center,draw=black, xshift=0.3cm](double compressed){$3 \times 5 \times 20$ DCT img\\manipulated and\\double compressed};
			
			\node[below=of manipulated, align=center, yshift=1.25cm] (simple) {Simple compression\\experiment};
			\node[below=of double compressed, align=center, yshift=1.25cm] (double) {Double compression\\experiment};
			
			\draw[arrow] (flickr.east) -- (dct images.west)
			node[midway,above] {$\mathcal{C}_1$}
			node[midway,below, align=center] {$\qf{1} \in \{60$,\\$75, 90\}$};
			
			\draw[arrow] (dct images.east) -- (pixel images.west)
			node[midway,above] {$\mathcal{D}_1$};
			\draw[arrow] (pixel images.east) -- (manipulated.west)
			node[midway,above] {Inpainting}
			node[midway,below] {BtB~\cite{bertazzini2024beyond}};
			\draw[arrow] (manipulated.east) -- (double compressed.west)
			node[midway,above] {$\mathcal{C}_2$}
			node[midway,below, align=center] {$\qf{2} \in \{50, 60$,\\$75, 90, 95\}$};	
			
			\draw[arrow] (manipulated.south) -- (simple.north);
			\draw[arrow] (double compressed.south) -- (double.north);
		\end{tikzpicture}
		\caption{Images forgery process for both simple and double compression experiments. $\mathcal{C}_1, \mathcal{D}_1, \mathcal{C}_2$ are the compressions and decompression defined in Fig.\ref{fig:scenarios} such as the quality factors $\qf{1}$ and $\qf{2}$.}
		
		\label{fig:dataset_flow}
	\end{figure*}
	
	\begin{figure*}[t]
		\centering
		\includegraphics[width=\textwidth]{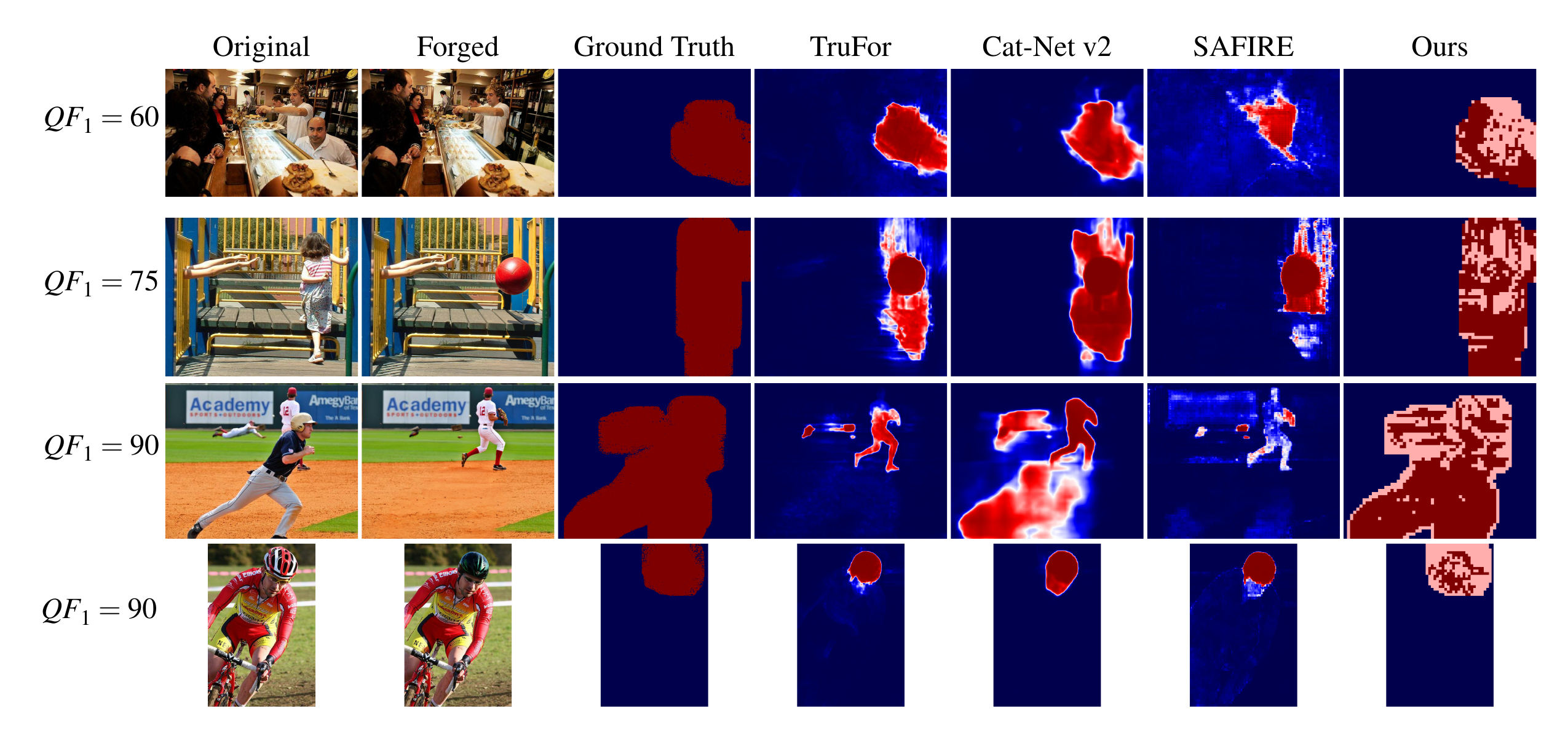}
		\caption{Visual comparison of inpainting localization. Original images have been compressed using $\qf{1} \in \{60,75,90,90\}$ from top to bottom. Then, they have been decompressed, manipulated and stored in PNG. Note that the masks and the predictions are at the pixel level and not at the block level like the metrics. In ours predictions, pink (resp. red) corresponds to unsolved blocks (resp. incompatible) but they are both classified as manipulated.}
		\label{fig:simple compression comparison}
	\end{figure*}

	\begin{figure*}[h!]
		\centering
		\includegraphics[width=\textwidth]{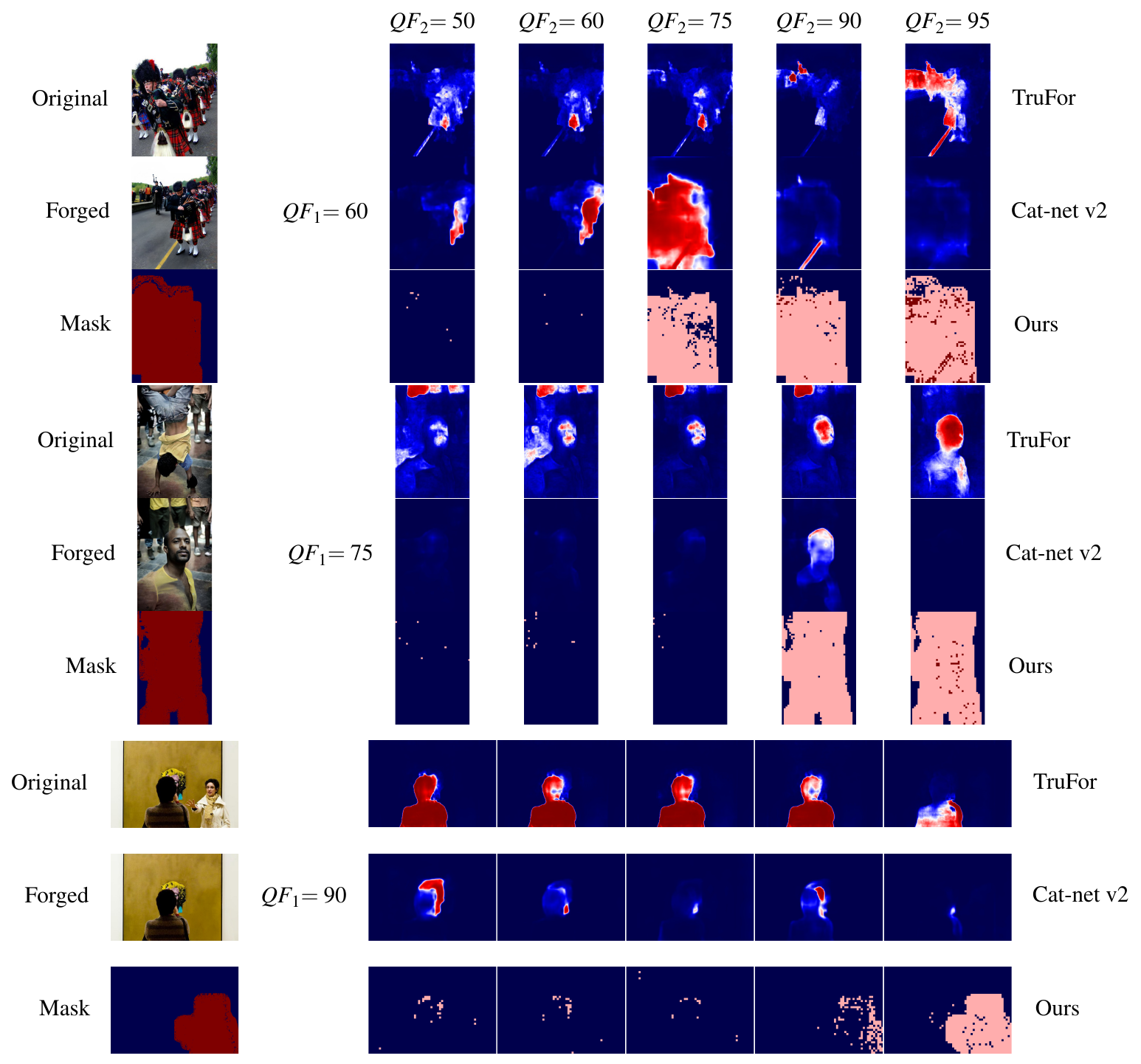}
		\caption{Localization results in double-compressed inpainted images. Each image has been compressed at $\qf{1}$, modified in the pixel domain, and recompressed at $\qf{2}$. The detection is done on the pixels values $\mathbf{F}$ for all models. In our predictions, pink corresponds to unsolved blocks, and red to incompatible blocks with $N=5000$ iterations. There are only few incompatible blocks because theoretical constraints in double compression are not tight enough to efficiently reduce the search space.}
		\label{fig:double compression comparison}
	\end{figure*}

	\subsection{Experimental setup}
	
	Because of our working assumptions we need to control the JPEG pipeline used to create manipulated images. Thus, we took the first 20 images from the Flickr30k dataset~\cite{young-etal-2014-image} and compressed and decompressed them at $\qf{1} \in \{60,75,90\}$ using a known JPEG pipeline to obtain $3\times20$ images. The DCT and IDCT algorithms used were \texttt{islow} from \texttt{libjpeg} with standard quantization tables. Then, we used a method called Beyond-the-Brush~\cite{bertazzini2024beyond} to manipulate all the 60 images (inpainting) without any human supervision. In the experiment, we consider the case where the manipulated images have not been compressed again after modification and the case where the image undergone a second compression at $\qf{2} \in \{50,60,75,90,95\}$ to obtain at the end $3\times5\times 20 = 300$ manipulated images. The flowchart  in Fig.~\ref{fig:dataset_flow} illustrates this dataset creation.


	The number of iterations of our algorithm is arbitrary fixed at $N=5000$. For most compatible blocks, a solution will be found in fewer iterations. However, for clipped blocks, with at least one pixel at 0 or one pixel at 255, the equations presented in section~\ref{sec:constraint} are no longer true because the rounding error can be bigger than $\frac{1}{2}$. Therefore, in presence of a clipped block, the algorithm will not apply the reduction of the search space which drastically increases the number of iterations.

	To compare our method, we use the TruFor model~\cite{guillaro2023trufor}, the Cat-Net v2 model~\cite{kwonLearningJPEGCompression2022a} and the SAFIRE~\cite{kwon2024safire} model. All of them are deep neural networks trained mainly on splicing and/or copy-move datasets, however, they are all based on artifact fingerprints and the JPEG compression applied before the modification should be sufficient to have two different areas in the image: one with the JPEG artifacts and one without. Moreover, Cat-Net v2 is specifically trained to detect JPEG artifacts using an RGB stream combined with a DCT stream. These three models output a probability map, so to obtain a binary map we use a threshold of 0.5 as proposed in Cat-Net v2 paper~\cite{kwonLearningJPEGCompression2022a}.

	The metrics used are all permuted metrics, this means that we take the best metric using the prediction or 1 minus the prediction. This is done to evaluate how well models can distinguish each region. Moreover, every metric is computed by blocks. To get the state of the block we apply the same rule to both the mask and the prediction: a block is classified as manipulated if at least one pixel of this block is classified as such. Finally, the two metrics used are the balanced accuracy, defined as:
	\begin{equation}
		ACC = \frac{TPR + TNR}{2},	
	\end{equation}
	and the False Positive Rate $FPR = FP/(FP + TN)$, where $TPR = TP/(TP+FN)$ denotes the True Positive Rate, $TNR = TN/(FP+TN)$ the True Negative Rate, $TP$ the True Positives, $FN$ the False Negatives, $FP$ the False Positives and $TN$ the True Negatives.

	\subsection{Results}
	
	\begin{table}[h]
		\centering
		{\renewcommand{\arraystretch}{1.25}
			\begin{tabular}{cC{2.25cm}C{2.25cm}}
				\multicolumn{1}{c|}{\textbf{}}  & Balanced Accuracy (ACC) & False Positive Rate (FPR) \\ \hline
				\multicolumn{1}{c|}{TruFor~\cite{guillaro2023trufor}}     & 73.62          & 0.21\\ 
				\multicolumn{1}{c|}{Cat-Net v2~\cite{kwonLearningJPEGCompression2022a}} & 75.54          & 0.48\\
				\multicolumn{1}{c|}{SAFIRE~\cite{kwon2024safire}} & 67.13          & 6.65 \\ 
				\multicolumn{1}{c|}{Ours}       & \textbf{99.98} & \textbf{0}
		\end{tabular}}
		
		\caption{Localization results on 60 images of size between $496\times 280$ and $496\times 496$. The balanced accuracy and the false positive rate are defined at the block level. Results has been averaged over $\qf{1} \in \{60,75,90\}$. There was no visible dependency on the $\qf{}$.}
		\label{tab: simple compression}
	\end{table}

	For the single compression setup, Fig.~\ref{fig:simple compression comparison} shows a visualization of the output of our method compared to other forgery localization methods, and Table~\ref{tab: simple compression} shows the metrics associated with this experiment. We can see that all models are relatively good for this task but our method is better, especially in terms of false positives. These results match the statistics obtained in the last section: all compatible blocks are detected as is and some incompatible blocks are also found. The pink blocks in our prediction represent unsolved blocks. Those are clipped blocks for which the theoretical upper bound can not be applied so the searching space is not reduced and the algorithm is not able to explore all the space.
	
	\begin{table*}[t]
		\centering
		\resizebox{\textwidth}{!}{{\renewcommand{\arraystretch}{1.25}
				\begin{tabular}{cl|ccccc|ccccc|ccccc}
					& \multicolumn{1}{c|}{QF1} & \multicolumn{5}{c|}{60}                                                                                   & \multicolumn{5}{c|}{75}                                                                                                 & \multicolumn{5}{c}{90}                                                                                                                        \\ \cline{2-17} 
					& \multicolumn{1}{c|}{QF2} & 50                          & 60                          & 75            & 90            & 95            & 50                          & 60                          & 75                          & 90            & 95            & 50                          & 60                          & 75                          & 90                                  & 95            \\ \hline
					\multicolumn{1}{c|}{}                         & ACC                      & 56.3                        & 55.4                        & 64.3          & 70.4          & 72.4          & 62.2                        & 58.6                        & 56.3                        & 71.6          & 71.6          & 57.8                        & 58.2                        & 56.8                        & 55.7                                & 68.8          \\[-0.1cm]
					\multicolumn{1}{c|}{\multirow{-2}{*}{TruFor~\cite{guillaro2023trufor}}} & FPR                      & \textbf{3.4}                & 2.6                         & 1.5           & 0.8           & 0.1           & \textbf{3.4}                & \textbf{1.8}                & \textbf{1.6}                & 0.8           & 0.1           & \textbf{4.0}                & 3.7                         & 2.6                         & 0.9                                 & 0.4           \\[0.15cm]
					\multicolumn{1}{c|}{}                         & ACC                      & 52.2                        & 53.6                        & 68.3          & 61.3          & 57.6          & 56.8                        & \textbf{67.4}               & 52.8                        & 64.6          & 60.1          & 53.0                        & 53.5                        & 56.5                        & 50.6                                & 63.5          \\[-0.1cm]
					\multicolumn{1}{c|}{\multirow{-2}{*}{Cat-Net v2~\cite{kwonLearningJPEGCompression2022a}}} & FPR                      & 4.0                         & 2.4                         & 1.3           & 0.5           & 0.5           & 13.0                        & 1.9                         & 5.9                         & 0.1           & 0.1           & 6.2                         & \textbf{3.1}                & \textbf{0.5}                & 2.6                                 & 0.0           \\[0.15cm]
					
					\multicolumn{1}{c|}{}                         & ACC                      & \textbf{59.6}               & \textbf{62.8}               & 61.9          & 64.0          & 65.7          & \textbf{62.6}               & 62.4                        & \textbf{58.9}               & 60.3          & 63.8          & \textbf{59.7}               & \textbf{63.0}               & \textbf{60.6}               & \textbf{60.2}                       & 63.5          \\[-0.1cm]
					\multicolumn{1}{c|}{\multirow{-2}{*}{SAFIRE~\cite{kwon2024safire}}} & FPR                      & 6.7                         & 19.8                        & 14.3          & 16.0          & 13.4          & 17.0                        & 17.3                        & 12.7                        & 12.2          & 2.3           & 22.2                        & 8.7                         & 13.7                        & 8.6                                 & 8.3           \\[0.15cm]
					\multicolumn{1}{c|}{}                         & ACC                      & {\color[HTML]{9B9B9B} 50.0} & {\color[HTML]{9B9B9B} 50.0} & \textbf{92.5} & \textbf{98.6} & \textbf{99.4} & {\color[HTML]{9B9B9B} 50.3} & {\color[HTML]{9B9B9B} 50.1} & {\color[HTML]{9B9B9B} 50.1} & \textbf{97.8} & \textbf{99.2} & {\color[HTML]{9B9B9B} 50.4} & {\color[HTML]{9B9B9B} 50.6} & {\color[HTML]{9B9B9B} 50.4} &  54.6         & \textbf{99.0} \\
					\multicolumn{1}{c|}{\multirow{-2}{*}{Ours}}   & FPR                      & {\color[HTML]{9B9B9B} 0.0}  & {\color[HTML]{9B9B9B} 0.0}  & \textbf{0.1}  & \textbf{0.0}  & \textbf{0.0}  & {\color[HTML]{9B9B9B} 0.3}  & {\color[HTML]{9B9B9B} 0.3}  & {\color[HTML]{9B9B9B} 0.0}  & \textbf{0.0}  & \textbf{0.0}  & {\color[HTML]{9B9B9B} 1.4}  & {\color[HTML]{9B9B9B} 1.0}  & {\color[HTML]{9B9B9B} 0.8}  &  \textbf{0.1} & \textbf{0.3} 
			\end{tabular}}
		}
		\caption{Localization results averaged over 20 images of size between $496\times280$ and $496\times496$. The balanced accuracy (ACC) and the false positive rate (FPR) are defined at the block level.}
		\label{tab:double compression block}
	\end{table*}

	For the double compression setup, the visual results are presented in Fig.~\ref{fig:double compression comparison}. As expected, when $\qf{1}$ is bigger than $\qf{2}$, almost all blocks are compatible (some clipped blocks are unsolved) and we are not able to detect any forgery. However, in the other case, when $\qf{1}$ is smaller than $\qf{2}$, the detection is very good compared to the state-of-the-art methods. Note that when $\qf{1} = \qf{2} > 85$ we are also able to detect parts of the forgery. Moreover, in a double compression setup, we can see that blocks are mostly either compatible or unsolved and there are almost no incompatible blocks. This is because the upper bound is not tight enough and is not able to reduce the searching space enough to bring it to a reasonable size that could be entirely explored in 5000 iterations.

	Table~\ref{tab:double compression block} show the block-wise metrics associated with this experiment. Again the metrics of our method match very closely to the statistics of the last section. In particular, Table~\ref{tab:double compression block} shows a FPR almost null at 5000 iterations (computed over more than 1M predictions). There is also no detection when $\qf{1}$ is bigger than $\qf{2}$ which was expected based on the last section statistics but note that, other state-of-the-art methods also have poor results in this setup. When $\qf{1}$ is smaller than $\qf{2}$, our method outperforms other methods with a very high detection power.

	\section{Conclusion and Perspectives}

	\subsection{Conclusion}

	This paper has presented a new approach to solving the problem of verifying the authenticity of an image under the assumption that they were JPEG compressed before a potential forgery. Our method is based on the notion of compatibility of JPEG blocks (existence of an antecedent) or incompatibility (absence of antecedent) which makes it very reliable under some assumptions (no False Positive). We extended this notion of compatibility to any number of compression, decompression or color transforms. We proposed a local search algorithm to look after the antecedent and combined it with a theoretical upper bound to reduce the search space.

	We have seen that the type of modification (inpainting, splicing, or copy-move) can create up to 3 types of mismatch (grid, quantization table, and pipeline mismatch). We then studied the link between mismatch type and block compatibility. In particular, we saw that when we look at the pixel image without recompression, almost all mismatched blocks will be incompatible and easy to detect. But when looking at the recompressed (or recompressed then decompressed) image, some modified blocks may remain compatible because of the quantization tables used.

	Finally, we conducted a forgery localization experiment to compare with state-of-the-art methods. It appears that without recompression, our method can accurately detect every modified block without error and thus outperforms other methods. If the image is recompressed after modification, our method can be very accurate when the second compression is weaker than the first one and also outperforms other methods. However, if the second compression is stronger than the first one, our method is unable to detect anything.

	\subsection{Perspectives}

	This work should be seen as a proof of concept of using JPEG compatibility to detect modified images. Indeed, this method requires strong assumptions to work properly. In particular, the exact JPEG pipeline of the original image must be known up to its DCT algorithm implementation. For future work, there are at least two axes to transform this method into an off-the-shelf model.

	The first axis is to make it more robust to unknowns. We could start to generalize it to unknown quantization tables by estimating a dictionary of potential tables and verifying the compatibility using each table. But the real challenge is to make it robust to any DCT (and IDCT) algorithm to be able to handle any image without knowing its main pipeline.
	The second axis is to improve the speed of the algorithm. For high $\qf{}$, the algorithm requires thousands of iterations for a single block to ensure no False Positive. This means that for big images, it can take hours to analyze them. The speed could be improved either by finding tighter upper bounds to reduce the search space or by finding a new algorithm to find antecedent (or prove incompatibility). In particular, we believe the upper bound could be improved with a probabilistic constraint on the neighborhood. This would create a trade-off between a controlled False Positive Rate and the algorithm efficiency.

	Finally, this method only relies on the property of non-surjectivity of compression and decompression processes. Such property could exist in other compressed formats. For example, forgery  H26x or HEIC formats could be the source of incompatibilities.

	\section*{Acknowledgments}
	\noindent The work presented in this paper received funding from the European Union’s Horizon 2020 research and innovation program under grant agreement No 101021687 (project “UNCOVER”).

	\bibliographystyle{plain}
	\bibliography{incompatibility}

\end{document}